\documentclass{article}

\usepackage{arxiv}

\usepackage[utf8]{inputenc} 
\usepackage[T1]{fontenc}    
\usepackage{hyperref}       
\usepackage{url}            
\usepackage{booktabs}       
\usepackage{amsfonts}       
\usepackage{nicefrac}       
\usepackage{microtype}      
\usepackage{lipsum}		
\usepackage{graphicx}
\usepackage{natbib}
\usepackage{doi}

\usepackage{graphicx}
\usepackage{amsmath}
\usepackage{amsfonts}
\usepackage{amssymb}
\usepackage{enumitem}
\usepackage{multirow}
\usepackage{makecell}
\usepackage[font=small,labelfont=bf,tableposition=top]{caption}
\DeclareCaptionLabelFormat{andtable}{#1~#2  \&  \tablename~\thetable}

\usepackage{subcaption}
\usepackage{tabularx}
\usepackage{hyperref}
\usepackage[nameinlink]{cleveref}
\usepackage{float}
\usepackage{tikz}
\usetikzlibrary{fit,positioning}
\usetikzlibrary{arrows}
\usepackage{algorithm}
\usepackage{algorithmicx}
\usepackage{algpseudocode}
\usepackage[colorinlistoftodos]{todonotes}

\newcommand{\bs}{\boldsymbol}

\Crefname{figure}{Fig.}{Figs.}
\crefname{figure}{fig.}{figs.}
\Crefname{equation}{Eq.}{Eqs.}
\crefname{equation}{eq.}{eqs.}

\DeclareSymbolFont{slant}{OT1}{\familydefault}{m}{sl}
\SetSymbolFont{slant}{bold}{OT1}{\familydefault}{b}{sl}
\DeclareSymbolFontAlphabet{\mathsl}{slant}

\title{Improving the Accuracy and Efficiency of Online Calibration for Simulation-based Dynamic Traffic Assignment}

\author{
	\href{https://orcid.org/0000-0003-3767-460X}{\includegraphics[scale=0.06]{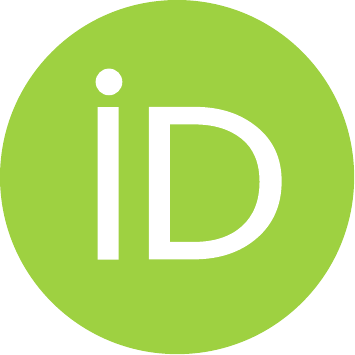}\hspace{1mm}Haizheng Zhang} \\
	Google LLC \\
	Mountain View, CA, 94043 \\
	\texttt{haizhengz@google.com} \\
	\And
	\href{https://orcid.org/0000-0002-9327-9455}{\includegraphics[scale=0.06]{orcid.pdf}\hspace{1mm}Ravi Seshadri} \\
	Technical University of Denmark \\
	Lyngby, Denmark \\
	\texttt{ravse@dtu.dk} \\
	\And
	{\hspace{1mm}A. Arun Prakash} \\
	Caliper Corporation \\
	Newton, MA, 02461 \\
	\texttt{arun@caliper.com} \\
	\And
	\href{https://orcid.org/0000-0003-0203-9542}{\includegraphics[scale=0.06]{orcid.pdf}\hspace{1mm}Constantinos Antoniou} \\
	Technical University of Munich \\
	Munich 80333, Germany \\
	\texttt{c.antoniou@tum.de} \\
	\And
	{\hspace{1mm}Francisco C. Pereira} \\
	Technical University of Denmark \\
	Lyngby, Denmark \\
	\texttt{camara@dtu.dk} \\
	\And
	\href{https://orcid.org/0000-0002-9635-9987}{\includegraphics[scale=0.06]{orcid.pdf}\hspace{1mm}Moshe Ben-Akiva} \\
	Massachusetts Institute of Technology \\
	Cambridge, MA, 02139 \\
	\texttt{mba@mit.edu} \\
}

\date{May 5, 2021}

\renewcommand{\headeright}{}
\renewcommand{\undertitle}{Accepted Manuscript \href{https://doi.org/10.1016/j.trc.2021.103195 }{doi: 10.1016/j.trc.2021.103195}}
\renewcommand{\shorttitle}{Improving the Accuracy and Efficiency of Online Calibration for Simulation-based DTA}

\hypersetup{
pdftitle={Improving the Accuracy and Efficiency of Online Calibration for Simulation-based Dynamic Traffic Assignment},
pdfsubject={eess.SY},
pdfauthor={Haizheng Zhang, Ravi Seshadri, A. Arun Prakash, Constantinos Antoniou, Francisco C. Pereira, Moshe Ben-Akiva},
pdfkeywords={Online Calibration, Dynamic Traffic Assignment, Simulation, Constrained Extended Kalman Filter},
}

\begin{document}
\maketitle

\begin{abstract}
	Simulation-based Dynamic Traffic Assignment models have important applications in real-time traffic management and control. The efficacy of these systems rests on the ability to generate accurate estimates and predictions of traffic states, which necessitates online calibration. A widely used solution approach for online calibration is the Extended Kalman Filter (EKF), which---although appealing in its flexibility to incorporate any class of parameters and measurements---poses several challenges with regard to calibration accuracy and scalability, especially in congested situations for large-scale networks. This paper addresses these issues in turn so as to improve the accuracy and efficiency of EKF-based online calibration approaches for large and congested networks. First, the concept of state augmentation is revisited to handle violations of the Markovian assumption typically implicit in online applications of the EKF. Second, a method based on graph-coloring is proposed to operationalize the partitioned finite-difference approach that enhances scalability of the gradient computations. 
	
	Several synthetic experiments and a real world case study demonstrate that application of the proposed approaches yields improvements in terms of both prediction accuracy and computational performance. The work has applications in real-world deployments of simulation-based dynamic traffic assignment systems.
\end{abstract}

\keywords{Online Calibration \and Dynamic Traffic Assignment \and Simulation \and Constrained Extended Kalman Filter}

\section{Introduction}
Traffic congestion is a pervasive problem that affects transportation networks worldwide and its severity continues to increase amidst the emergence of disruptive technologies such as ride-hailing. In the United States, during peak hours in 2016, trips took 35\% more time on average than during non-peak hours (compared to 20\% in 2010). According to \citet{series2016fhwa}, the average duration of congestion in US traffic systems was 4.7 hours daily in 2016, compared with 4.3 hours in 2009 \citep{taylor20102009}. Apart from increases in travel time and delays, congestion exacerbates air pollution, energy consumption and emissions. 
Congestion incurred an estimated \$160 billion annual cost for extra time and fuel in 2014, and the cost was expected to be \$192 billion in 2020 \citep{schrank2015urban}. 

The adverse impacts of congestion have led to an increasing emphasis on the development of tools for traffic management, to alleviate congestion by more efficiently utilizing existing infrastructure. Effective traffic management necessitates the generation of accurate short-term predictions of traffic states and in this context, simulation-based Dynamic Traffic Assignment (DTA) systems have gained prominence over the years. A key component of these real-time DTA systems is online calibration which attempts to adjust simulation parameters in real-time to
match as closely as possible simulated measurements with real-time surveillance data. 

A widely used solution approach for online calibration is the Extended Kalman Filter (EKF), which is appealing due to its flexibility to incorporate any class of parameters and measurements. However, the EKF poses several challenges with regard to accuracy and efficiency. Specifically, the applicability to congested scenarios and scalability to large-scale networks hinder the real-world deployment of EKF. This constitutes the key motivation underlying our paper.
On one hand, accuracy in the context of large real-world networks---especially under congestion---is a concern due to the violations of the Markovian assumption typically implicit in online applications of the EKF. On the other hand, computational performance is a key challenge with the EKF---which involves repeated computations of the Jacobian matrix (or the gradient of the DTA simulator)---as traditional finite-difference techniques do not scale well with the size of the parameter space. 

With these motivating considerations, the broad objectives of this paper are to first, propose techniques to enhance the accuracy and efficiency of EKF-based online calibration approaches, and second, to test the proposed approaches on synthetic and real-world networks. 

This paper contributes to the existing literature on online calibration in three respects. First, the concept of state augmentation is revisited to handle violations of the Markovian assumption typically implicit in online applications of the EKF. Second, a method based on graph-coloring is proposed to operationalize the partitioned finite-difference approach that enhances scalability of the gradient computations. Finally, synthetic experiments and a real world case study demonstrate that application of the proposed approaches yields improvements in terms of both prediction accuracy and computational performance. The work has important applications in real-world deployments of simulation-based dynamic traffic assignment systems.

The remainder of this paper is structured as follows. Section \ref{sec:LitRev} reviews relevant literature on online calibration. An overview of the state-space model and the EKF solution approach is provided in Section \ref{sec:Background_OC}.
Section \ref{sec:StateAug} revisits the concept of state-augmentation and Section \ref{sec:PartSP} proposes an approach to improve computational performance. A case-study on the network of expressways in Singapore is presented in Section \ref{sec:CaseStudy} and Section \ref{sec:Conclusions} provides concluding remarks and directions for future research.

\section{Review of Literature}
\label{sec:LitRev}

The calibration of simulation-based DTA model systems has been extensively studied in two contexts: offline and online. The offline calibration problem involves determining historical values of demand and supply parameters of the simulator so as to best replicate historical network performance on an average day. In particular, the dynamic Origin-Destination (OD) estimation problem has been widely studied (for the static version see \citep{cascetta1984estimation,bell1983estimation,maher1983inferences}) using primarily Generalized Least Squares (GLS) approaches \citep{cascetta1993dynamic} and state-space modeling approaches \citep{ashok1996estimation,okutani1984dynamic}. The GLS-based optimization reformulations are typically under-determined problems as the number of non-zero demand flows is usually significantly larger than number of measurements, which has been addressed by using information from a prior or seed OD matrix. More recently, \cite{cascetta2013quasi} proposed a quasi-dynamic GLS estimator (based on the assumption that OD shares are constant across a reference period, whereas total flows leaving each origin vary for sub-periods within the reference period) to address the under-determined nature of the problem. Issues of computational scalability are addressed in \cite{osorio2019high} who propose a metamodel simulation-based optimization approach for the dynamic OD estimation problem. In contrast with the aforementioned approaches which focus on demand parameters, \cite{balakrishna2007offline} proposed 
a more generic solution method for the
offline problem based on the Simultaneous Perturbation
Stochastic Approximation (SPSA) algorithm that simultaneously incorporates demand and supply side parameters and is applicable to any type of measurement data. Several variants of the SPSA algorithm have since been proposed including weighted SPSA or W-SPSA \citep{lu2015enhanced}, cluster-wise SPSA or c-SPSA \citep{tympakianaki2015c}, discrete W-SPSA \citep{oh2019demand}, PC-SPSA \citep{qurashi2019pc}, and gradient approximation \citep{cipriani2011gradient}. A more detailed review of the offline calibration problem can be found in \cite{osorio2019high,djukic2014dynamic}.

The online calibration problem, in contrast with the offline version, involves recursively updating simulation parameters (such as OD-flows on the demand side and segment capacities or traffic dynamic parameters on the supply side) in real-time to best replicate current or prevailing traffic conditions. Existing online calibration approaches can be broadly categorized into those that focus on demand parameters, supply parameters and a combination of both types.

Early work on online calibration focused on demand-side parameters, specifically, the problem of dynamic OD estimation and prediction for real-time applications. \cite{ashok1996estimation} and \citep{ashok2000alternative} generalized the approach of \cite{okutani1984dynamic} and proposed a state-space model and a Kalman filter solution approach. The evolution of within-day OD flows was modeled using an autoregressive process and the problem was formulated in terms of deviations of OD flows from their historical values, so as to incorporate \textit{a priori} structural information. Along similar lines,      
\cite{zhou2007structural} applied a Kalman filtering solution approach, where the transition equation is
a polynomial trend filter that captures historical trends and structural deviations. Existing research has also tackled computational issues related to OD estimation and prediction: its performance on large-scale networks was addressed by \cite{bierlaire2004efficient}, and more recently by \cite{cipriani2011gradient} and \cite{cantelmo2015improving}; constraints on the state variables (such as non-negativity of OD flows) were explicitly handled in \cite{zhang2017improved}; stochasticity in the assignment matrix was incorporated in \cite{ashok2002estimation}.
More recently, \cite{marzano2018kalman} apply the extended Kalman filter to the quasi-dynamic estimation and updating of OD flows, relying on the quasi-dynamic assumption noted earlier. The authors propose a closed-form linearization of the measurement equation assuming an uncongested network and error-free assignment matrix. Trip-chaining within the state space model is considered explicitly in \cite{cantelmo2020incorporating} and once again, solved using a Kalman filter. Efficient online estimators are also proposed considering measurements other than just traffic counts (for instance, bluetooth data and turning volume data) in \cite{barcelo2010travel}, \cite{barcelo2013kalman} and \cite{lu2015kalman}.

With regard to the calibration of supply-side parameters,
\cite{zhou2002dynamic} proposed a dynamic programming
approach that uses macroscopic model approximations of the simulator to adjust flow propagation. \cite{antoniou2007nonlinear,antoniou2004line} extended the state-space model of \cite{ashok1996estimation} for the simultaneous online calibation of demand and supply parameters and proposed the use of the extended, limiting, iterated and unscented Kalman filters. \cite{hashemi2015integrated} also addressed the joint calibration of demand and supply parameters, and proposed an approach where the OD demands are calibrated using a least squares approach and supply-side parameters are calibrated using a feedback controller. 

A key challenge that arises in using the Extend Kalman Filter or EKF \citep{antoniou2007nonlinear} for online
applications is the linearization step that involves calculation of numerical derivatives of the simulator, which is computationally intensive, particularly when the number of calibration parameters is large. Several studies have attempted to address this issue. \cite{antoniou2004line,antoniou2007nonlinear} proposed the limiting EKF, which utilizes offline computed derivatives and drastically reduces the computational complexity of the EKF, without sacrificing significantly on accuracy in some settings. More recently, a dynamic Bayesian networks' view of state augmentation was investigated in \citep{zhang2018towards} to relax the underlying Markovian assumption of EKF. In regard to computational efficiency, dimensionality reduction techniques \citep{djukic2012application,prakash2017reducing,prakash2018improving}, and network decomposition and partitioning approaches \citep{frederix2014dynamic,huang2010algorithmic} have been applied to improve scalability of online calibration solution approaches.

In summary, despite the extensive research on online calibration, several issues have received limited attention with regards to EKF-based approaches. First, all online applications of the Kalman filter employ a Markovian assumption for analytical and computational tractability, which can adversely affect estimation and prediction accuracy in the case of sparse sensor availability---this is especially common on large networks or during congestion---an issue that has not received sufficient attention in the literature. Second, there has been insufficient work on operationalizing the partitioned finite-difference approach proposed in \citep{huang2010algorithmic} which can significantly improve the computational performance of gradient computations based on finite difference (FD). Finally, this work attempts to address these limitations pertaining to EKF-based online calibration approaches in the literature so as to aid real-world deployments of online simulation-based DTA systems. 

\section{Background: Online Calibration}
\label{sec:Background_OC}
The online calibration problem involves updating historical DTA model parameters (such as origin-destination (OD) flows on the demand side and segment capacities and traffic dynamics parameters on the supply side) in real-time to match simulated and current traffic conditions as closely as possible \citep{antoniou2004line}. The online calibration problem is typically formulated as a state space model which is briefly reviewed in this section. For more details, the reader is referred to \cite{antoniou2004line} and \cite{zhang2018online}. 

\subsection{The State Space Model}
The state space model is a classical approach that models dynamical systems by describing the probabilistic dependence between latent state variables and observed measurements. The state-space formulation consists of three main components: (i) a state vector that succinctly characterizes the system, (ii) a transition equation that captures the evolution of the system (in terms of the state vector) over time, and (iii) a measurement equation that captures the relationship between the state vector and the measurements or observations of the system.

Let $\boldsymbol x_h$ denote the state vector (OD flows and supply parameters to be calibrated) in time interval $h$, where $h \in\mathcal{H}=\{1,2,...,H\}$, and $\mathcal H$ is the set of discrete time intervals within the simulation period. The state space model can be formulated as,
\begin{align}
	\boldsymbol x_h = \boldsymbol f(\boldsymbol x_{h-1}, ..., \boldsymbol x_{h-p})+\boldsymbol w_h 
	\label{eq:abstract_trans} \\
	\boldsymbol{M}_h =\boldsymbol g(\boldsymbol x_h, ..., \boldsymbol x_{h-q+1})+\boldsymbol v_h
	\label{eq:abstract_meas}
\end{align}
where, \Cref{eq:abstract_trans} is the transition equation, and \Cref{eq:abstract_meas} is the measurement equation; $\bs f(\cdot)$ and $\bs g(\cdot)$ are functions that determine the transition and measurement relations; $p$ denotes the number of previous intervals' states that influence the current interval's state (depending on structural patterns as well as the time discretization of OD flows); 
$q$ denotes the number of previous states that influence the measurements in the current interval; $\boldsymbol{M}_h$ denotes the vector of measurements/observations in interval $h$, and $\bs w_h,\bs v_h$ are vectors of zero-mean noises that each follows a multivariate normal distribution.

Further, we assume the transition equation in \Cref{eq:abstract_trans} is modeled using a linear autoregressive process, and replace the generic function $\bs g(\cdot)$ in \Cref{eq:abstract_meas} with the simulator (DTA model) $\bs S(\cdot)$. 
Thus, the state space model is given by, 
\begin{align}
	\boldsymbol x_h = \sum_{k=h-p}^{h-1}\boldsymbol F^{k}_h\boldsymbol x_{k}+\boldsymbol w_h
	\label{eq:trans}
\end{align}
\vspace{-0.75 cm}
\begin{align}
	\boldsymbol{M}_h =\boldsymbol S(\boldsymbol x_h, ..., \boldsymbol x_{h-q+1})+\boldsymbol v_h
	\label{eq:meas}
\end{align}
where, $\boldsymbol F^{k}_h$ is a square matrix, representing the effect of $\boldsymbol x_{k}$ on $\boldsymbol x_{h}$.

As described in \cite{ashok1996estimation}, the state vector can be expressed in terms of deviations from historical values. This formulation has the advantage of incorporating \textit{a priori} structural information (e.g. spatial-temporal patterns of demand flows) in the calibration process. The state space model expressed in terms of deviations can be written as,
\begin{align}
	\Delta \boldsymbol x_h = \sum_{k=h-p}^{h-1}\boldsymbol F^{k}_h \Delta\boldsymbol x_{k}+\boldsymbol w_h
	\label{eq:abstract_trans_dev} 
\end{align}
\vspace{-0.75 cm}
\begin{align}
	\Delta \boldsymbol{M}_h =\boldsymbol S(\Delta \boldsymbol x_h + \boldsymbol x_h^H , ..., \Delta \boldsymbol x_{h-q+1} + \boldsymbol x_{h-q+1}^H ) - \boldsymbol S(\boldsymbol x_h^H , ..., \boldsymbol x_{h-q+1}^H )+\boldsymbol v_h
	\label{eq:abstract_meas_dev}
\end{align}
where $\Delta \boldsymbol x_h = \boldsymbol x_h - \boldsymbol x_h^H$, $\Delta \boldsymbol{M}_h = \boldsymbol{M}_h - \boldsymbol{M}_h^H$; $ \boldsymbol x_h^H $ and $\boldsymbol{M}_h^H$
represent historical values of the state and measurement vectors respectively in interval h. With a slight abuse of notation, we continue to use $\boldsymbol w_h , \boldsymbol v_h$ to represent the error terms in the transition and measurement equations. The error terms $\boldsymbol w_h , \boldsymbol v_h$ are assumed to be zero mean, uncorrelated with each other and uncorrelated across time intervals. Further the covariance matrices of $\boldsymbol w_h , \boldsymbol v_h$ are denoted by $\boldsymbol Q_h$ and $\boldsymbol R_h$ respectively.




\subsection{Solution Approach: Extended Kalman Filter}\label{sec:EKF}
The standard approach to solve the state space model defined previously is the Extended Kalman Filter or EKF \citep{antoniou2007nonlinear}. The EKF linearizes the non-linear measurement equation around the \textit{a priori} estimates and adopts the classical Kalman Filter procedure for linear state space models to estimate the state vector. Although the EKF does not guarantee optimality (in terms of minimizing the mean squared error), it has been shown to yield good results in practice. The EKF solution approach is briefly summarized here for completeness. Refer \cite{antoniou2007nonlinear} for more details on the EKF and \cite{zhang2017improved} for details on the constrained EKF which handles constraints on the state vector (such as non-negativity of OD flows).

In the description of the EKF that follows, the state space model is represented more compactly as follows, 
\begin{align}
	\boldsymbol X_h = \boldsymbol \Phi_{h-1}\boldsymbol X_{h-1}+\boldsymbol W_h \label{eq:SAt}
\end{align}
\begin{align}
	\Delta\boldsymbol{M}_h =\bs S\left(\boldsymbol X_h + \boldsymbol{\mathsl{x}}_h^H \right) - \bs S \left( \boldsymbol{\mathsl{x}}_h^H \right) +\boldsymbol V_h
\end{align}
where, the dependence of the state vector on multiple preceding intervals is handled using the concept of state augmentation \citep{ashok1996estimation,okutani1984dynamic}, by defining
\begin{align}
	\bs X_h =& \left[\Delta \bs x_h^\top, \Delta  \bs x_{h-1}^\top, ..., \Delta \bs x_{h-r+1}^\top\right]^\top \label{eq:State_aug_SV} \\
	\boldsymbol{\mathsl{x}}_h^H =& \left[(\bs x_h^H)^\top, (\bs x_{h-1}^H)^\top, ..., (\bs x_{h-r+1}^H)^\top\right]^\top \\
	\bs{\Phi}_{h} =& \left[\begin{array}{cc}
		\multicolumn{2}{c}{\begin{array}{cccc}
				\bs F_{h}^{h-1}& \bs F_{h}^{h-2}& \cdots& \bs F_{h}^{h-r}\end{array}}\\
		\bs I_{(r-1)n\times(r-1)n} & \bs 0_{(r-1)n\times n}
	\end{array}\right] 
	\label{eq:State_aug_trans}
\end{align}\label{eq:State_aug}

In the equations above, $r = \max\{p,q\}$ and $n$ is the number of DTA parameters to be calibrated in each interval (dimensionality of $\boldsymbol x_h$ in \Cref{eq:abstract_trans}). 
Note that in online applications, due to computational constraints, an approximation is typically used wherein the state variables from previous intervals $ \Delta  \bs x_{h-1}, ..., \Delta \bs x_{h-r+1} $ are not re-estimated in interval $h$. In other words, in interval $h$ \textit{only} $ \Delta \bs x_h $ is estimated \citep{ashok1996estimation}.   

With these definitions, the EKF algorithm is summarized below. 
The EKF algorithm involves three main steps: (i) time-update, (ii) linearization, and (iii) measurement update. 

In the time-update step, the \textit{a priori} estimates of state vector and covariance matrix (denoted $\hat{\boldsymbol{X}}_{h|h-1}$ and $ \boldsymbol{P}_{h|h-1} $) for the current interval are computed using the transition equation and the optimal estimates from the previous interval ($\hat{\boldsymbol{X}}_{h-1|h-1} $ and $\boldsymbol{P}_{h-1|h-1}$). 

In the second step, the measurement equation is linearized around the \textit{a priori} estimates of the current interval (e.g., $\hat{\boldsymbol{X}}_{h|h-1}$), which yields the following linearized measurement equation:
\begin{align}
	\Delta\boldsymbol{M}_h =\bs S\left(\hat{\boldsymbol{X}}_{h|h-1} + \boldsymbol{\mathsl{x}}_h^H \right) - \bs S \left( \boldsymbol{\mathsl{x}}_h^H \right) + \bs\Theta_h \bs X_h - \bs\Theta_h \hat{\boldsymbol{X}}_{h|h-1} +\boldsymbol V_h
\end{align}
where $ \bs\Theta_h = \nabla_{\boldsymbol{X_h}} \boldsymbol S( \boldsymbol{X_h}+ \boldsymbol{\mathsl{x}}_h^H )  \vert _{\hat{\boldsymbol{X}}_{h|h-1}} $ represents the Jacobian or gradient of the simulator with respect to $\bs X_h$ evaluated at $\hat{\boldsymbol{X}}_{h|h-1}$. Note that $\bs\Theta_h $ may also be expressed as,
\begin{align}
	\bs\Theta_h = \left[\bs H_h^h, \bs H_h^{h-1}, ..., \bs H_h^{h-r+1}\right] 
	\label{eq:Grad_EKF}
\end{align} 
where $ \bs H_h^k $ represents the gradient of the simulator in interval $h$ with respect to the parameter vector $ \Delta \bs x_k $.  

In the third step, the \textit{a priori} estimates are updated using the linearized measurement equation and the matrix ${{\boldsymbol{\Theta}_{h}}}$ to obtain the \textit{a posteriori} estimates of the state vector $\hat{\boldsymbol{X}}_{h|h}$ and its covariance matrix $\boldsymbol{P}_{h|h} $. In some settings, it may be necessary to impose constraints on the state vector in which case a constrained version of the EKF (termed C-EKF) may be applied \citep{zhang2017improved}. 

\begin{algorithm}[H]
	\begin{algorithmic}[]
		\caption{Extended Kalman Filter} \label{alg:EKF}
		\State Initialize
		\begin{align}
			\hat{\boldsymbol X}_{0|0} =\boldsymbol X_0 \label{eq:x0}\\
			\boldsymbol P_{0|0} =\boldsymbol P_0 \label{eq:P0}
		\end{align}
		\For{$h$ = $1$ to $H$}
		\State \textbf{Time Update}
		\State \textit{A priori} estimate
		\begin{align}
			\hat{\boldsymbol{X}}_{h|h-1} = \boldsymbol \Phi_{h-1}\hat{\boldsymbol{X}}_{h-1|h-1} \label{eq:xpred}  
		\end{align}
		\begin{align}
			\boldsymbol{P}_{h|h-1} =  {{\boldsymbol{\Phi}_{h-1}}} \boldsymbol{P}_{h-1|h-1}{ {\boldsymbol{\Phi}_{h-1}^\top}} + \boldsymbol{Q}_{h} \label{eq:Ppred} 
		\end{align}	
		\State \textbf{Linearization}
		\begin{align}
			\bs\Theta_h = \nabla_{\boldsymbol{X_h}} \boldsymbol S( \boldsymbol{X_h}+ \boldsymbol{\mathsl{x}}_h^H )  \vert _{\hat{\boldsymbol{X}}_{h|h-1}}   \label{eq:H}
		\end{align}		
		\State \textbf{Measurement Update}		
		\State Near-optimal Kalman gain 
		\begin{align}
			\boldsymbol{K}_{h} = \boldsymbol{P}_{h|h-1}{{\boldsymbol{\Theta}_{h}^\top}}\left({{\boldsymbol{\Theta}_{h}}}\boldsymbol{P}_{h|h-1}{{\boldsymbol{\Theta}_{h}^\top}} + \boldsymbol{R}_{h}\right)^{-1} \label{eq:Kg}
		\end{align}
		\State \textit{A posteriori} estimate
		\begin{align}
			\hat{\boldsymbol{X}}_{h|h} = \hat{\boldsymbol{X}}_{h|h-1} + \boldsymbol{K}_{h}\left(\boldsymbol{M}_{h} - \boldsymbol {g}_h(\hat{\boldsymbol{X}}_{h|h-1})\right) \label{eq:xpost} 
		\end{align}
		\begin{align}
			\boldsymbol{P}_{h|h} = \boldsymbol{P}_{h|h-1} - \boldsymbol{K}_{h} {{\boldsymbol{\Theta}_{h}}}\boldsymbol{P}_{h|h-1} \label{eq:Ppost}
		\end{align}
		\EndFor
	\end{algorithmic}
\end{algorithm}

\subsection{Challenges for Deployment}
Several challenges remain in the operationalization of the EKF based solution approaches for the online calibration problem, each of which is addressed in subsequent sections. These include (i) accurate parameter estimations in cases of less observability, and (ii) real-time computation performance. Both aspects are important to applications to large-scale and congested networks, which has been recognized as an essential challenge in DTA deployment \citep{peeta2001foundations}.


Estimating less observable DTA parameters is crucial when dealing with congestion scenarios and large-scale networks. During congestion, traffic flow sensors often become insensitive to the immediate changes on demand, making it only observable after considerable delays. On large networks, there are more trips that have longer travel times. They are measured by sensor measurements at different time intervals, usually with significant delays. Traffic estimation of those trips is hard and affects the predictive power of the DTA model. The issue of delayed measurements relating to less observable parameters is addressed in Section \ref{sec:StateAug} where the concept of state augmentation is revisited. 

The real-time performance is addressed in Section \ref{sec:PartSP}. In Kalman filter based DTA applications, the linearization of the measurement equation is a critical bottleneck that hinders computational performance since it requires computation of the Jacobian matrix. Traditional finite difference techniques do not scale well with the dimension of the parameter space, requiring $2n$ runs of the simulator (in each time interval) for a parameter vector of dimension $n$. This issue generally prevents DTA models from being scalable to large networks.

\section{State Augmentation}\label{sec:StateAug}
In this section, we revisit the concept of state augmentation briefly introduced in Section \ref{sec:EKF}. As noted previously, for the computational tractability of online applications, an approximation is typically used: state variables from previous intervals (\Cref{eq:State_aug}) are not re-estimated in the current interval. However, this approximation may significantly affect accuracy, depending on the location of sensors on the network and the spatio-temporal distribution of demand.  

\subsection{Probabilistic Representation}
\begin{figure}[hbt!]
	\centering
	\resizebox{280pt}{90pt}{
		\begin{tikzpicture}
			\tikzstyle{main}=[circle, minimum size = 13mm, thick, draw =black!80, node distance = 16mm]
			\tikzstyle{ellipsis}=[circle, minimum size = 13mm, thick, draw =white!100, node distance = 16mm]
			\tikzstyle{connect}=[-latex, thick]
			\tikzstyle{box}=[rectangle, draw=black!100]
			\node[main, fill = white!100] (x1) [label=center:$\boldsymbol x_1$] { };
			\node[main] (x2) [right=of x1,label=center:$\boldsymbol x_2$] {};
			\node[main] (x3) [right=of x2,label=center:$\boldsymbol x_3$] {};
			\node[ellipsis] (x35) [right=of x3,label=center:$\boldsymbol \cdots$] {};
			\node[main] (x4) [right=of x35,label=center:$\boldsymbol x_H$] {};
			\node[main, fill = black!10] (M1) [below=of x1,label=center:$\boldsymbol M_1$] { };
			\node[main, fill = black!10] (M2) [below=of x2,label=center:$\boldsymbol M_2$] { };
			\node[main, fill = black!10] (M3) [below=of x3,label=center:$\boldsymbol M_3$] { };
			\node[main, fill = black!10] (M4) [below=of x4,label=center:$\boldsymbol M_H$] { };
			
			\path (x1) edge [connect] (x2)
			(x2) edge [connect] (x3)
			(x3) edge [connect] (x35)
			(x35) edge [connect] (x4)
			(x1) edge [connect] (M1)
			(x2) edge [connect] (M2)
			(x3) edge [connect] (M3)
			(x4) edge [connect] (M4);
		\end{tikzpicture}
	}
	\caption{State space model with measurements}
	\label{fig:markovmodel}
\end{figure}
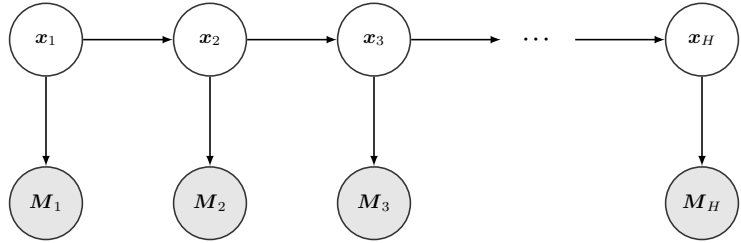

In order to examine the implications of the above approximation, consider the following Dynamic Bayesian Network (DBN) representation of the state space model (\Cref{fig:markovmodel}). The shaded nodes are observed measurements; the unshaded ones are latent state variables which cannot be directly measured.

From a DBN perspective, \Cref{fig:markovmodel} exhibits a probabilistic directed graphical model structure that defines the factors of the joint probability, the directed edges depict conditional probability with connected nodes being random variables. Specifically,
\begin{align}
	f(\bs x_{1:H}, \bs M_{1:H})=f(\bs x_1)f(\bs M_1|\bs x_1)\prod_{i=1}^{H-1}f(\bs x_{i+1}|\bs x_i)f(\bs M_{i+1}|\bs x_{i+1})
\end{align}
The representation in \Cref{fig:markovmodel} shows that for example, $\boldsymbol x_2$ uniquely determines $\boldsymbol x_3$. In other words, conditioned on $\boldsymbol x_2$, $\boldsymbol x_1$ and $\boldsymbol x_3$ are independent. Similarly, when $\boldsymbol{x}_2$ is given, $\bs x_1$ does not affect $\boldsymbol{M}_2$. This is the Markovian/memoryless assumption in state space models and is implicit in the typical online calibration process. 
More specifically in Kalman filtering, the prior $\bs x_{h|h-1}$ is given by the transition equation from $\bs x_{h-1|h-1}$. Upon observing $\bs M_h$, we obtain the posterior estimator $\bs x_{h|h}$, which is further used to construct the prior $\bs x_{h+1|h}$ as the process continues. As we can see, it does not update any previous states (e.g., $\bs x_{h|h+i}$) based on future observations (e.g., $\bs M_{h+i}$). This makes the online estimation fast, as we reduce the complexity of the parameter space from $\boldsymbol{x}_{1:h}$ to $\boldsymbol{x}_h$ for each time slice $h$.

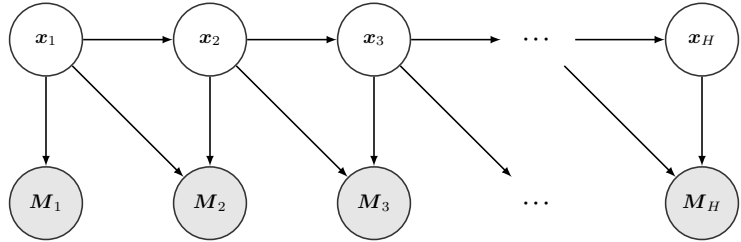
\begin{figure}[hbt!]
	\centering
	\resizebox{280pt}{90pt}{
		\begin{tikzpicture}
			\tikzstyle{main}=[circle, minimum size = 13mm, thick, draw =black!80, node distance = 16mm]
			\tikzstyle{connect}=[-latex, thick]
			\tikzstyle{ellipsis}=[circle, minimum size = 13mm, thick, draw =white!100, node distance = 16mm]
			\tikzstyle{box}=[rectangle, draw=black!100]
			\node[main, fill = white!100] (x1) [label=center:$\boldsymbol x_1$] { };
			\node[main] (x2) [right=of x1,label=center:$\boldsymbol x_2$] {};
			\node[main] (x3) [right=of x2,label=center:$\boldsymbol x_3$] {};
			\node[ellipsis] (x35) [right=of x3,label=center:$\boldsymbol \cdots$] {};
			\node[ellipsis] (M35) [below=of x35, label=center:$\boldsymbol \cdots$] {};
			\node[main] (x4) [right=of x35,label=center:$\boldsymbol x_H$] {};
			\node[main, fill = black!10] (M1) [below=of x1,label=center:$\boldsymbol M_1$] { };
			\node[main, fill = black!10] (M2) [below=of x2,label=center:$\boldsymbol M_2$] { };
			\node[main, fill = black!10] (M3) [below=of x3,label=center:$\boldsymbol M_3$] { };
			\node[main, fill = black!10] (M4) [below=of x4,label=center:$\boldsymbol M_H$] { };
			\path (x1) edge [connect] (x2)
			(x1) edge [connect] (M2)
			(x2) edge [connect] (x3)
			(x2) edge [connect] (M3)
			(x3) edge [connect] (M35)
			(x3) edge [connect] (x35)
			(x35) edge [connect] (x4)
			(x1) edge [connect] (M1)
			(x2) edge [connect] (M2)
			(x3) edge [connect] (M3)
			(x35) edge [connect] (M4)
			(x4) edge [connect] (M4);
		\end{tikzpicture}
	}
	\caption{A DBN with measurement equation contradicting the Markovian assumption}
	\label{fig:non-markovmodel}
\end{figure}

Although the inference task is simplified with the Markovian assumption, it may be problematic when we have a delayed system. Consider a case where the $i$th element of $\boldsymbol{x}_h$---denoted by $\boldsymbol{x}_h(i)$---only has an impact on measurement $\boldsymbol{M}_{h+1}$ in time slice $h+1$. Clearly, it is impossible to accurately estimate $\boldsymbol{x}_{h}(i)$ with standard Kalman filtering techniques when we only know $\boldsymbol{M}_{h}$. We illustrate this example more intuitively with its corresponding DBN representation in \Cref{fig:non-markovmodel}. 
Noticeably the Markovian assumption no longer holds. Applying the Kalman filter on this example essentially ignores the true diagonal relations.

\subsection{Example with the Toy Network}\label{sec:toy_ex}
The violation of the Markovian assumption in \Cref{fig:non-markovmodel} has implications on system observability, which is illustrated using the simple example network in \Cref{fig:fullyobservedex} for an OD estimation problem (estimating dynamic OD demands from traffic flow measurements). This network has two OD pairs and $s1$ and $s2$ are two flow-count sensors that report aggregated flow within each 5-minute time interval.

\begin{minipage}{\textwidth}
	\vspace*{20pt}
	\begin{minipage}[b]{0.55\textwidth}
		\centering
		\includegraphics[width=\linewidth]{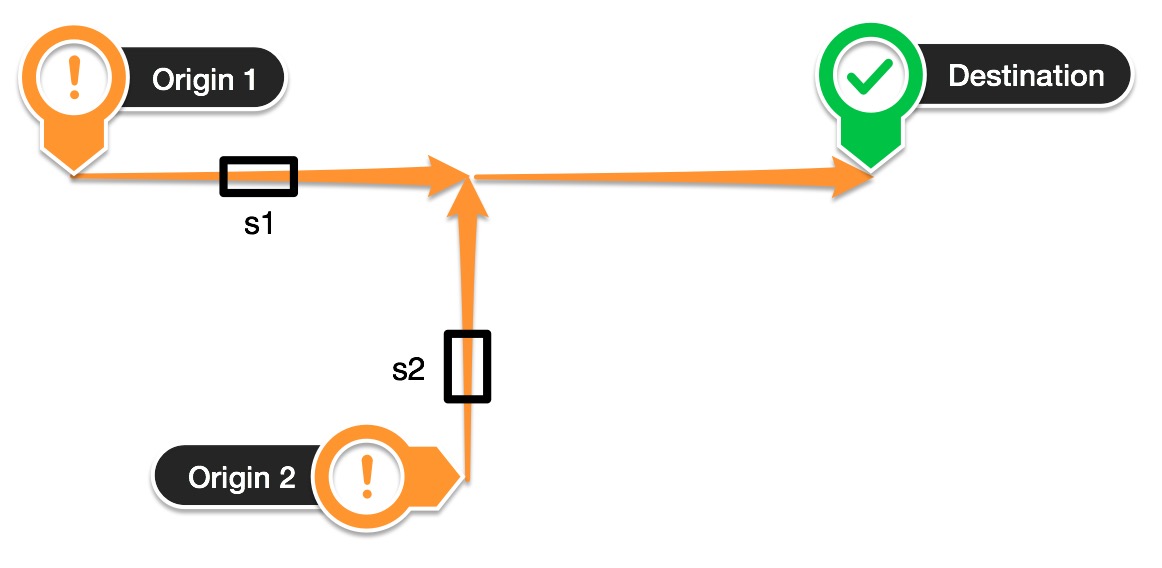}
		\captionof{figure}{A road network example and the sensor placement that ensures no delay in capturing the states}
		\label{fig:fullyobservedex}
	\end{minipage}
	\hspace{\fill}
	\begin{minipage}[b]{0.37\textwidth}
		\captionof{table}{Example OD and sensor flows for toy network}
		\centering
		\begin{tabular}{ccc}
			\Xhline{1pt}
			& \textbf{t=1} & \textbf{t=2} \\ \hline
			$\mathbf{O_1}\mathbf{D}$ & 30  & 24  \\
			$\mathbf{O_2D}$ & 20   & 18  \\ \hline
			\textbf{s1}      & 30  & 24  \\
			\textbf{s2}      & 20   & 18  \\ \Xhline{1pt}
		\end{tabular}
		\vspace{10pt}
		\label{table:fully_obs}
	\end{minipage}
	\vspace*{20pt}
\end{minipage}

In this example, we make three assumptions: (1) each link takes 1 time interval to traverse, (2) all vehicles will travel the same distance within each interval, meaning a sensor either captures all or nothing from an OD pair in each interval, and (3) there is no measurement error in sensor flow counts.

\Cref{table:fully_obs} shows an example of the OD and sensor flows in two intervals. Note $s1$ only captures $O_1D$ in the same interval and $s2$ captures $O_2D$. The OD flow inference is instant: we can read off measurements as OD flows. In this case, the system has no time-delay and the state space model in \Cref{fig:markovmodel} is accurate.

Next, we introduce a delay in measuring the OD: we change the sensor placement scheme as shown in \Cref{fig:observedlagex}. The measurements for $s2$ and $s3$ are listed in \Cref{tab:delay_obs}. The key change is that now $s3$ captures ${O_1D}$ and $O_2D$ with a delay of one interval. This introduces correlation between states and measurements across time intervals, making the Markovian assumption invalid. We can still read off $s2$ to estimate $O_2D$, but we have no information about $O_1D$ at $t=1$ unless we also know $s3$ at $t=2$. Thus, failing to model the correlation across intervals could lead to no update for some hidden states thereby reducing calibration accuracy.

\begin{minipage}{\textwidth}
	\vspace*{20pt}
	\begin{minipage}[b]{0.55\textwidth}
		\centering
		\includegraphics[width=\linewidth]{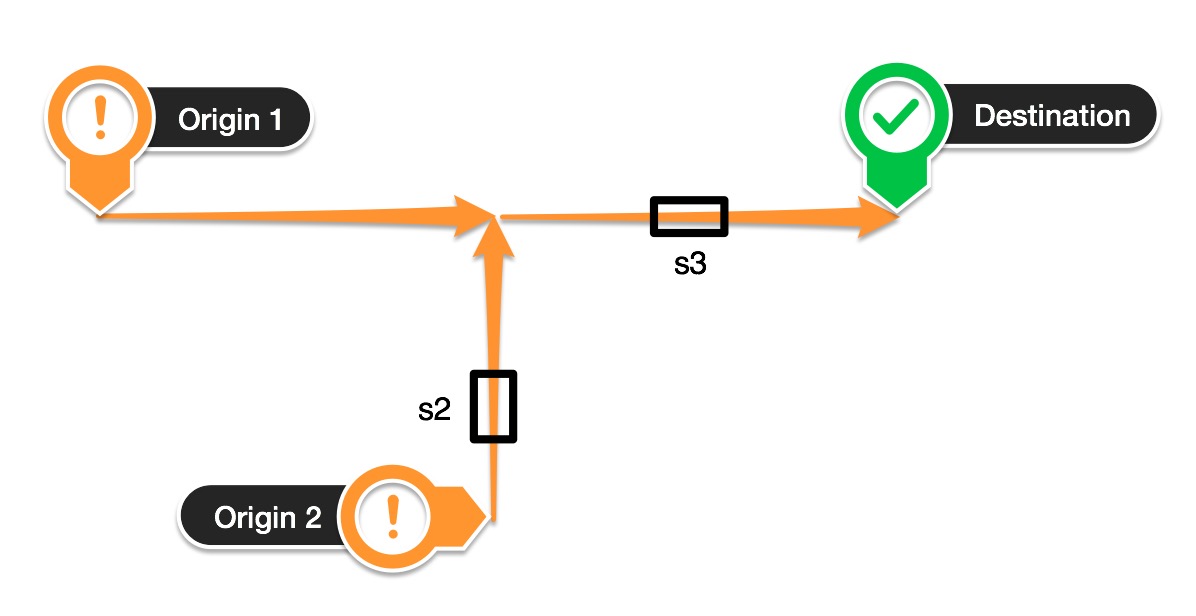}
		\captionof{figure}{The toy network and a sensor placement scheme with lag between OD and flow counts}
		\label{fig:observedlagex}
	\end{minipage}
	\hspace{\fill}
	\begin{minipage}[b]{0.37\textwidth}
		\captionof{table}{Example OD estimation with lag in measurements}
		\centering
		\vspace*{10pt}
		\begin{tabular}{ccc}	
			\Xhline{1pt}
			& \textbf{t=1} & \textbf{t=2} \\ \hline
			$\mathbf{O_1}\mathbf{D}$ & ?  & ?  \\
			$\mathbf{O_2D}$ & 20   & 18  \\ \hline
			\textbf{s2}      & 20   & 18  \\ 
			\textbf{s3}      & 0  & 50  \\ \Xhline{1pt}
		\end{tabular}
		\vspace*{10pt}
		\label{tab:delay_obs}
	\end{minipage}
	\vspace*{20pt}
\end{minipage}

\subsection{State Augmentation in DTA Models}
As noted previously, the process of state augmentation can mitigate the issue of hidden states. Recall \Cref{eq:State_aug_SV}, where the degree of augmentation is defined to be $r=\textnormal{max}\lbrace p,q \rbrace$. In principle, $q$ can be determined from a distribution of trip travel times on the network, ideally covering a majority of them (either from simulated data or observed data of trip times where detailed knowledge of the spatio-temporal patterns of congestion may not be required). However, in practice, $q$ may be unnecessary large resulting in excessively large computational time. Thus it may be beneficial use a relatively small $q$ and maintain system observability to a large extent. This depends on a combination of factors including the spatio-temporal distribution of demand and congestion, network topology and sensor coverage. In the case study in Section \ref{sec:CaseStudy}, we discuss a simple heuristic to set the degree of augmentation.   

\begin{figure}[hbt!]
	\centering
	\includegraphics[width=0.8\linewidth]{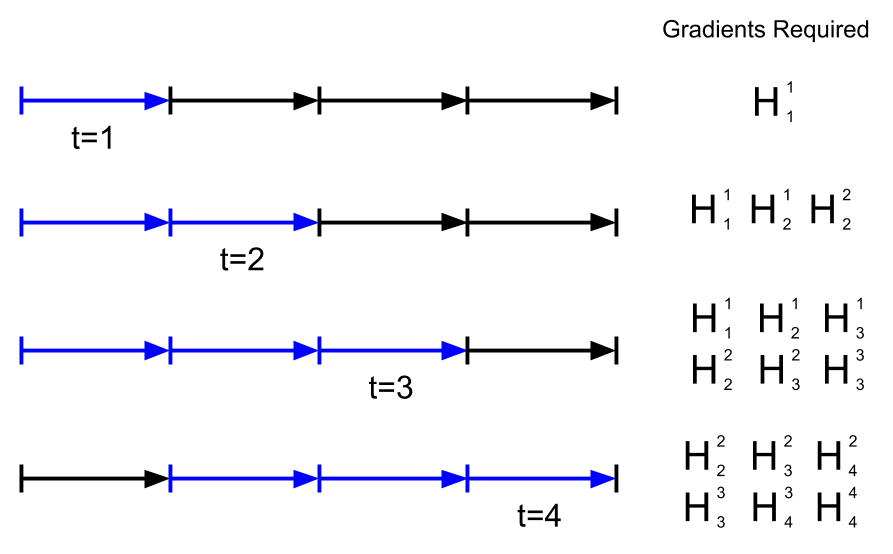}
	\caption{The intervals (blue) to calibrate state vectors on and the gradients needed as time progresses for a state augmentation with a degree of 3}
	\label{fig:interval}
\end{figure}

\Cref{fig:interval} illustrates the online calibration process across time intervals (denoted by $t$) assuming a state augmentation degree of 3 ($r$ in \Cref{eq:State_aug_SV}). The last blue interval marked ``$t=$" in each row is the current simulation time interval. In the first row, the measurement $\boldsymbol M_1$ is available and used to calibrate the first state vector $\boldsymbol{x}_1$. Starting from the second interval, the state augmentation allows the calibration of $\boldsymbol{x}_1$ and $\boldsymbol{x}_2$ with measurements $\boldsymbol M_1$ and $\boldsymbol M_2$, which specifically requires $\bs H_2^{1}$ in \Cref{eq:Grad_EKF}. In the 3rd row, we additionally need $\bs H_2^{1}$, $\bs H_3^{1}$ and $\bs H_3^{2}$ compared to the non-augmented model. Noticeably, in the 3rd row of \Cref{fig:interval}, we have 6 gradients to estimate. In general there are $O(r!)$ matrices to estimate for each interval, where $r$ is the degree of augmentation. 

\subsubsection{Staggered Horizons for Gradient Estimation}
\label{ssec:staggered_horizons}
To decrease the computational cost, we assume that gradients do not change significantly when revisited. Thus, we can skip re-estimating the same gradient. There are two benefits to this:
\begin{enumerate}
	\item We can reuse estimated gradients e.g., $\bs H_1^{1}$ in the 2nd and 3rd row in \Cref{fig:interval}.
	\item We can estimate some gradients upfront when they can be computed easily and use them later.
\end{enumerate}

Regarding the validity of the above assumption, in the case of recurring congestion, if the modelling of the temporal variation in OD flows---as represented by the transition equation---is reasonably accurate, we can be assured that the gradients are relatively stable. However, in the presence of non-recurrent events where the gradients may change significantly when revisited, the implications of this assumption require a systematic investigation. Alternatively, one could switch to recomputing the gradients in the presence of these special events and incidents.

\begin{figure}[hbt!]
	\centering
	\includegraphics[width=0.8\linewidth]{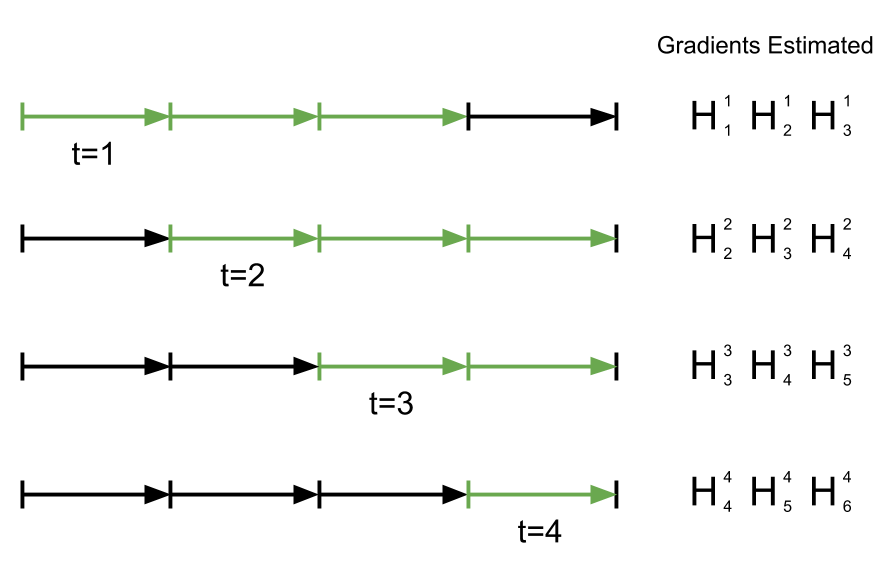}
	\caption{The estimated gradients and the intervals (green) calculated upon for state vector at t=k with ``Staggered Horizons''}
	\label{fig:staggered}
\end{figure}

Regardless, with this assumption, we can simplify the gradient estimation procedure. Here we present the ``Staggered Horizons" method in \Cref{fig:staggered}. At interval $t=1$, we measure the impact of $\bs x_1$ on the current and 2 future intervals, resulting in $\bs H_1^1$, $\bs H_2^1$ and $\bs H_3^1$. Similarly we estimate 3 more gradients for $t=2$ and $t=3$. By the time we finish estimating gradients for interval $t=k$, we can perform calibration for the same interval as we have all the gradients needed ($k$th row in \Cref{fig:interval}). 

In terms of computational cost, when using finite differences, we only need to perturb state vector $\bs x_t$ for interval $t$ (same as no state augmentation) and run the continuous simulation for intervals $[t, t+r-1]$ ($r$ times longer than the non-augmented model). Thus, the overall computational cost is $r$ times greater compared with the non-augmented model, which is significantly more efficient than $r!$ without using this technique. There is an additional benefit in only perturbing one state vector $\bs x_t$, which relates to the overhead when starting simulations. Specifically, running the continuous simulation for $t=1,2,3$ is faster than running them separately. By applying staggered horizons, we skip the overhead associated with starting new simulations to perturb $\bs x_{t+1}$, etc. 

\subsubsection{Cost with the Constrained EKF}
\label{ssec:cekf_cost}
Another increased cost is from the optimization step in the Constrained EKF \citep{zhang2017improved}. This step essentially involves solving a quadratic program with constraints. Now, with $r$ times more variables to estimate, the optimization is computationally more expensive. In the case study, we show that the computation time for the optimization increases by a factor of roughly 30 when $r=3$, but is still small compared to the time spent on the gradient estimation.

\subsubsection{Simulator Requirements and State Augmentation in Operation}

Enabling state augmentation also needs a few features of the simulator because it runs the same interval $r$ times via evaluation, as shown in \Cref{fig:interval}. The simulator should be able to:
\begin{enumerate}
	\item store a snapshot of the current traffic state (e.g., locations of simulated vehicles and any unassigned demand).
	\item resume simulation from a traffic snapshot.
\end{enumerate}

With these two features, now we can demonstrate the state estimation process in \Cref{fig:interval}.
\begin{enumerate}
	\item Current time $t=1$, simulation starts. We store a snapshot $s_1$ at the beginning of $t=1$ and estimate state vector $\bs x_1$.
	\item Current time $t=2$, simulation resumes from $s_1$. We store a snapshot $s_2$ at the beginning of $t=1$,  re-estimate $\bs x_1$, and estimate $\bs x_2$.
	\item Current time $t=3$, simulation resumes from $s_2$. We re-estimate $\bs x_1$ for the last time, store a snapshot $s_3$ at the beginning of $t=2$, re-estimate $\bs x_2$, and estimate $\bs x_3$.
	\item Current time $t=4$, simulation resumes from $s_3$. We re-estimate $\bs x_2$ for the last time, store a snapshot $s_4$ at the beginning of $t=3$, re-estimate $\bs x_3$ and estimate $\bs x_4$. 
	\item The process continues as described above.
\end{enumerate}

\subsection{Performance on the Synthetic Network}
The following sections demonstrate the performance of state augmentation on a synthetic network, typically in a congested scenario. 

\subsubsection{Road network}
In order to address the issues described in Section \ref{sec:toy_ex}, the EKF solution procedure (described in Algorithm \ref{alg:EKF}) should use the augmented states (as per \Crefrange{eq:State_aug_SV}{eq:State_aug_trans}). Applying state augmentation allows the previous estimated states to be adjusted based on the latest measurements. We will demonstrate the impact of this on estimation and prediction accuracy with the following example.

\begin{figure}[hbt!]
	\centering
	\includegraphics[width=1.03\linewidth]{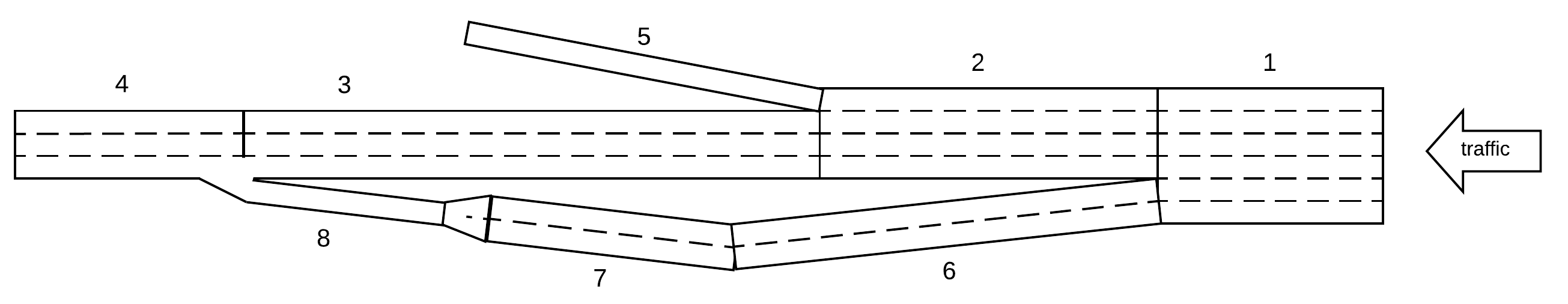}
	\caption{Synthetic network}
	\label{fig:toynet}
\end{figure}

The simulations are conducted on the synthetic network shown in \Cref{fig:toynet}, which consists of 8 segments and 2 origin-destination (OD) demand pairs.   
Each segment has a sensor that captures the mean speed and the aggregated flow for each specified time interval. We present network attributes in \Cref{tbl:toy_stats_new}, including segment lengths, free flow speeds and free flow travel times. We also show the mainstream and off-ramp OD flow statistics in \Cref{tbl:toy_demand} for the simulation period of 14:00-19:00. Under free flow conditions (\Cref{tbl:toy_stats_new}), the main stream OD travel time is 76 and 84 seconds. 

\begin{table}[hbt!]
	\centering
	\caption{Specifications of each segment on the synthetic network}
	\footnotesize
	\label{tbl:toy_stats_new}
	\begin{tabular}{ccccccccc}
		\Xhline{1pt} 
		Segment ID                   & 1     & 2     & 3     & 4     & 5     & 6     & 7     & 8     \\ \hline
		Length (meter) & 297.5 & 553.8 & 493.1 & 351.2 & 408.6 & 666.7 & 377.3 & 183.0 \\
		\begin{tabular}[c]{@{}c@{}}Free flow \\speed (mph)\end{tabular} & 50    & 50    & 50    & 50    & 20    & 50    & 50    & 50    \\
		\begin{tabular}[c]{@{}c@{}}Free flow travel\\time (second)\end{tabular} & 13.31  & 24.8  & 22.1  & 15.7  & 45.7  & 29.8  & 16.9  & 8.19 \\ \Xhline{1pt}
	\end{tabular}
\end{table}

\subsubsection{Data generation}
Once again, we consider the online OD estimation problem. The data we need are the aggregated flow counts for 5-minute intervals within 14:00-19:00. We obtain the flows by running the simulation with the given demand, whose major statistics are shown in \Cref{tbl:toy_demand}. Meanwhile, the supply parameters do not change during the simulation.

\begin{table}[hbt!]
	\centering
	\caption{Demand statistics for simulation period 14:00-19:00}
	\footnotesize
	\label{tbl:toy_demand}
	\begin{tabular}{ccccccc}
		\Xhline{1pt}
		\multirow{2}{*}{OD pair} & \multicolumn{5}{c}{OD flows at percentile (veh/hour)}   & \multirow{2}{*}{Mean OD flow (veh/hour)} \\
		& 10\% & 25\% & 50\% & 75\% & 90\% &                       \\ \hline
		Mainstream               & 3670 & 3882 & 4086 & 4446 & 4940 & 4220                  \\
		Off-ramp                 & 0    & 168  & 336  & 480  & 708  & 350                  \\ \Xhline{1pt}
	\end{tabular}
\end{table}

\begin{figure}[H]
	\centering
	\includegraphics[width=0.8\linewidth]{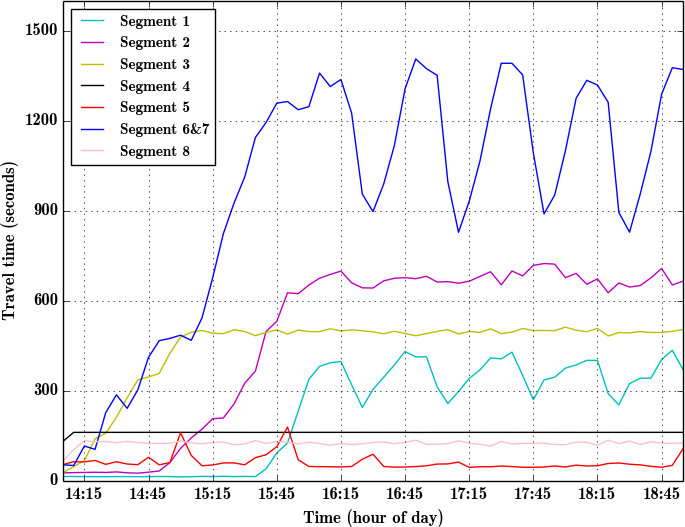}
	\caption{Link travel times on the toy network with the modified supply parameters}
	\label{fig:LinkTTbyInterval}
\end{figure}

When assigning the demand to the network, we obtained the true congested link travel times in \Cref{fig:LinkTTbyInterval}. The scenario is heavily congested: it is evident that the congestion from Segment 4 propagates backwards to Segment 8 and 3, then to other upstream segments. The oscillation in Segment 6, 7 and 2 is because of the stop and go traffic conditions. The key condition for the Markovian assumption is whether there is any OD flow that cannot be inferred from the sensor counts in the same interval. Starting at 16:00, the traversal time of Segment 1 exceeds 5 minutes and traffic flow only passes sensors on Segment 2 in the subsequent interval. Thus it is impossible to distinguish the two OD flows from only the sensor counts on Segment 1. Note that even with the help of sensors on Segment 2 and 6, we still cannot estimate the OD flows with certainty. Thus, the violation of the Markovian assumption will start at around 16:00. 

When setting up the experiment, we also need a transition equation that describes how the 2 OD flows evolve. Since the true time-dependent demand is known, we can obtain a true auto-regression (AR) process to model the transition equation (\Cref{eq:SAt}). Based on the Akaike information criterion (AIC), the best model was found to be an AR(5) model which is hereafter used for the transition equation. The historical OD demands are constructed by suitably perturbing the true OD demands. 

\subsubsection{Experiments and results}
The performance of the Constrained EKF (CEKF, \citet{zhang2017improved}) algorithm is now compared using three different degrees of state augmentation: 
\begin{itemize}[]
	\item CEKF(1): CEKF with original state space model (no augmentation), AR(5) transition model
	\item CEKF(2): CEKF with 2nd-order augmented state space model, AR(5) transition model 
	\item CEKF(5): CEKF with 5th-order augmented state space model, AR(5) transition model 
\end{itemize}

Note that the degree of augmentation can be lower than the transition AR degree (as the Experiment 1 and 2 above), because by convention we can apply the following approximation. When determining $\Phi_{h-1}$ in \Cref{eq:Ppred}, we use its corresponding upper left sub-matrix of the true transition matrix in the original AR process. The whole transition matrix (corresponding to 5 previous states) is still used to obtain the best prior in \Cref{eq:trans}.

\begin{table}[H]
	\centering
	\caption{RMSN for OD flow estimation and predictions for 15:00-19:00}
	\label{tbl:SArmsn}
	\begin{tabular}{ccccc}
		\Xhline{1pt}
		\multirow{2}{*}{Experiment} & \multirow{2}{*}{\begin{tabular}[c]{@{}c@{}}Estimation\\ RMSN\end{tabular}} & \multicolumn{3}{c}{Prediction RMSN} \\
		&    & Step 1     & Step 2     & Step 3    \\ \hline
		CEKF(1)  & 13.5\%      & 21.0\%     & 26.2\%     & 34.7\%    \\
		CEKF(2)  & 9.8\%    & 18.8\%     & 24.2\%     & 31.9\%    \\ 
		CEKF(5)  & 10.8\%     & 15.4\%     & 19.3\%  & 26.6\%  \\ \Xhline{1pt}
	\end{tabular}
\end{table}

After running the online calibration experiments above, we obtain the results in \Cref{tbl:SArmsn}, which illustrates the performance of the three models with the same AR(5) transition equation for predictions. For state estimation, CEKF(2) and CEKF(5) have smaller errors than CEKF(1), while CEKF(2) has the best estimation accuracy. However, in terms of prediction performance, CEKF(5) significantly outperforms CEKF(2), which is better than CEKF(1). This is likely because the CEKF(5) model estimates OD flows more accurately in the congested scenario after 16:00 (see \Cref{fig:LinkTTbyInterval}), underscoring the significant improvements obtainable through state augmentation in certain settings. 

\begin{figure}[H]
	\centering
	\makebox[\textwidth]{%
		\begin{minipage}[t]{0.33\textwidth}
			\includegraphics[width=\textwidth]{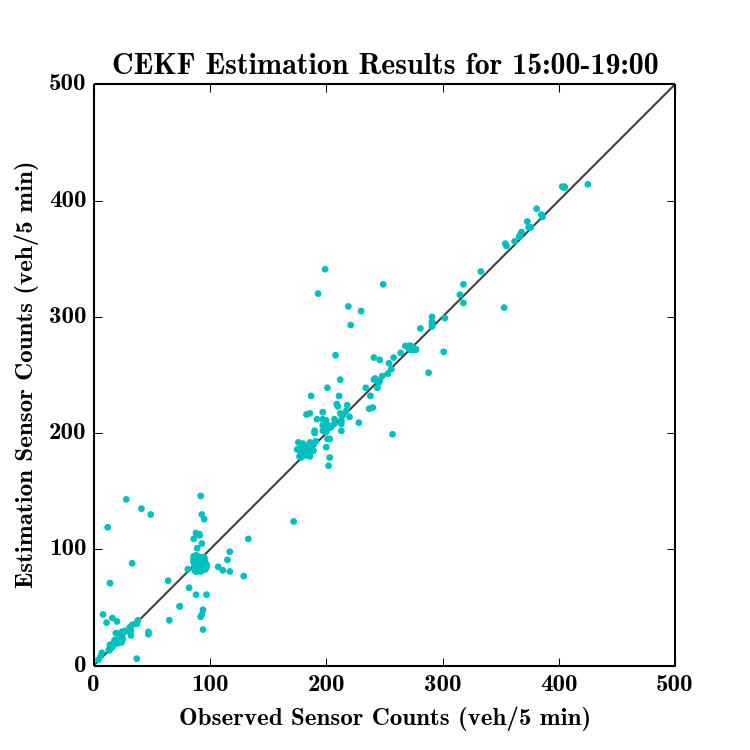}
			\includegraphics[width=\textwidth]{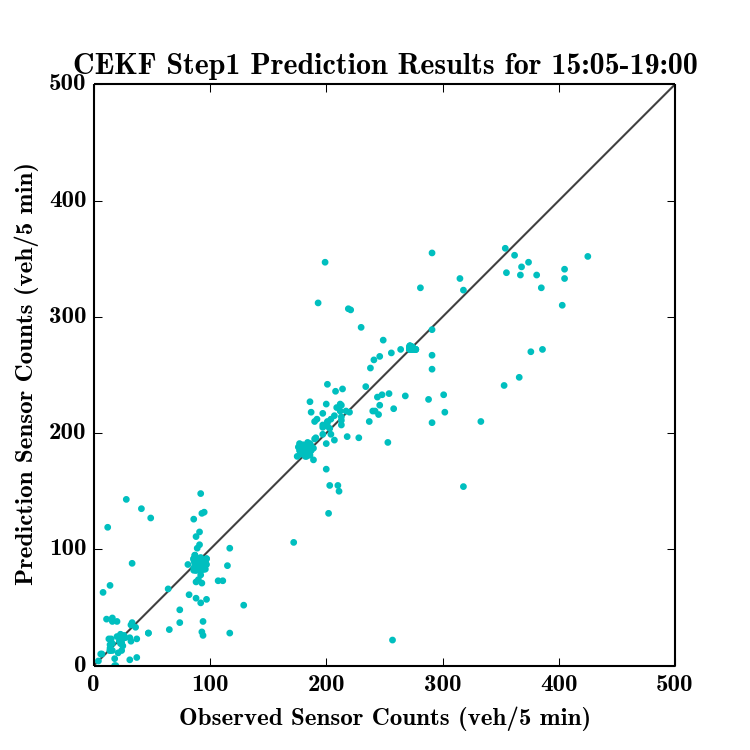}
			\includegraphics[width=\linewidth]{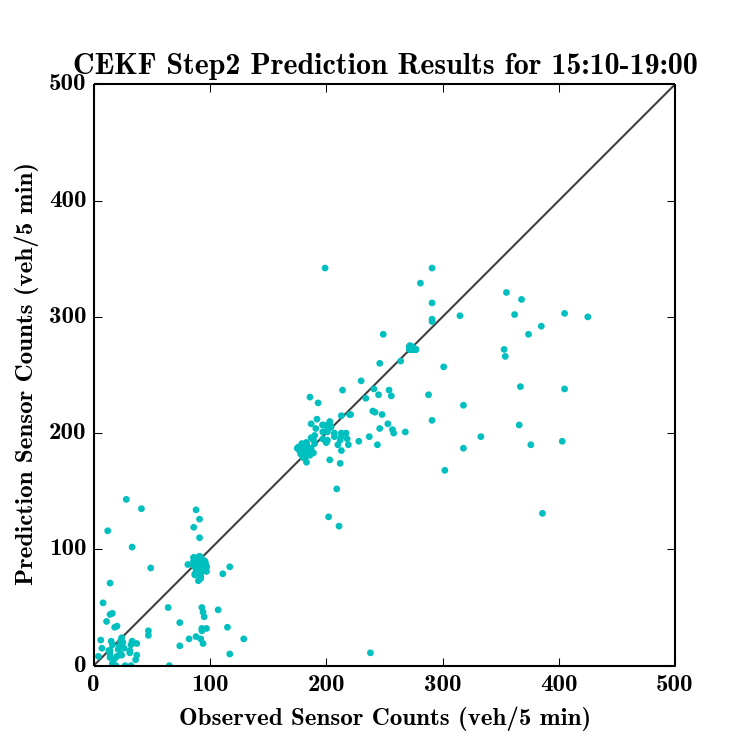}
			\includegraphics[width=\linewidth]{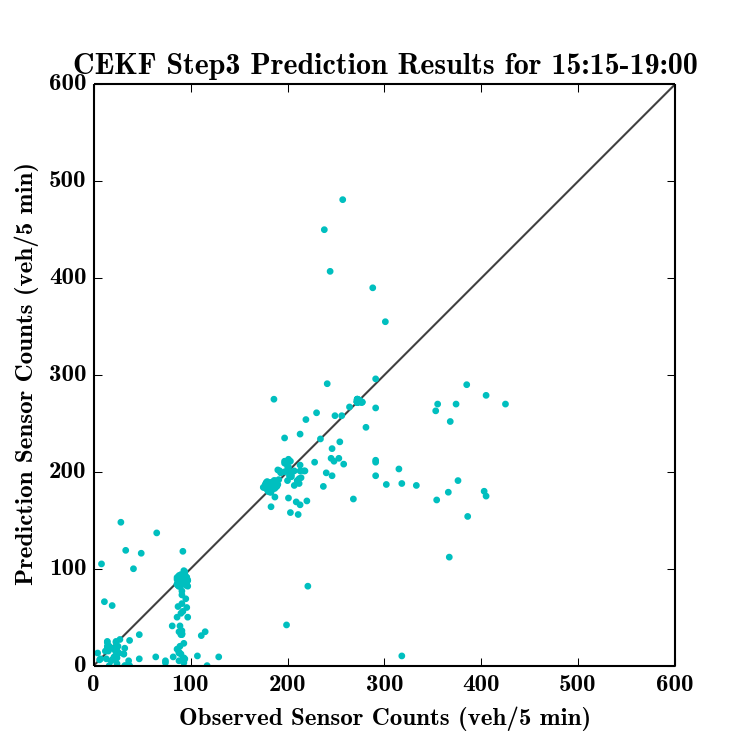}
		\end{minipage}%
		\begin{minipage}[t]{0.33\textwidth}
			\includegraphics[width=\textwidth]{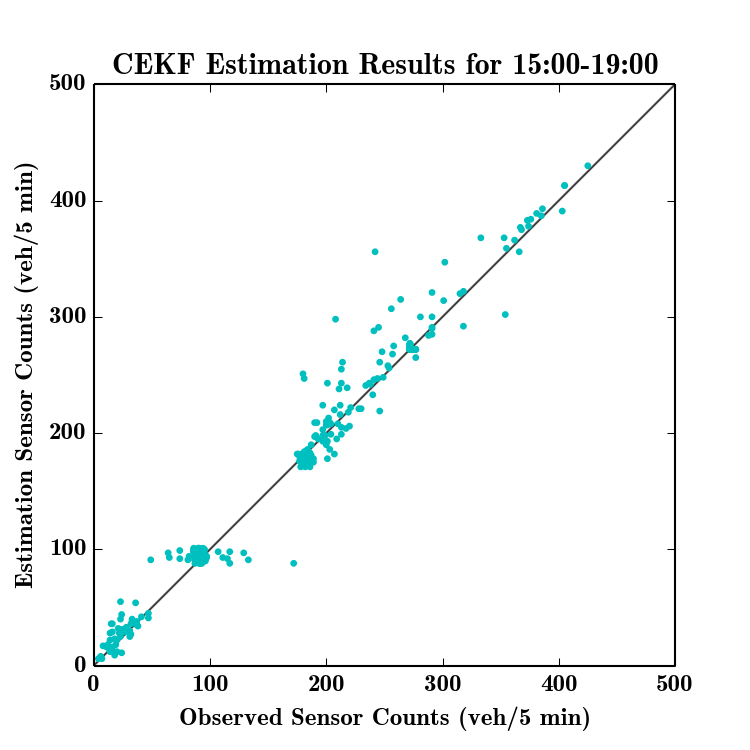}
			\includegraphics[width=\textwidth]{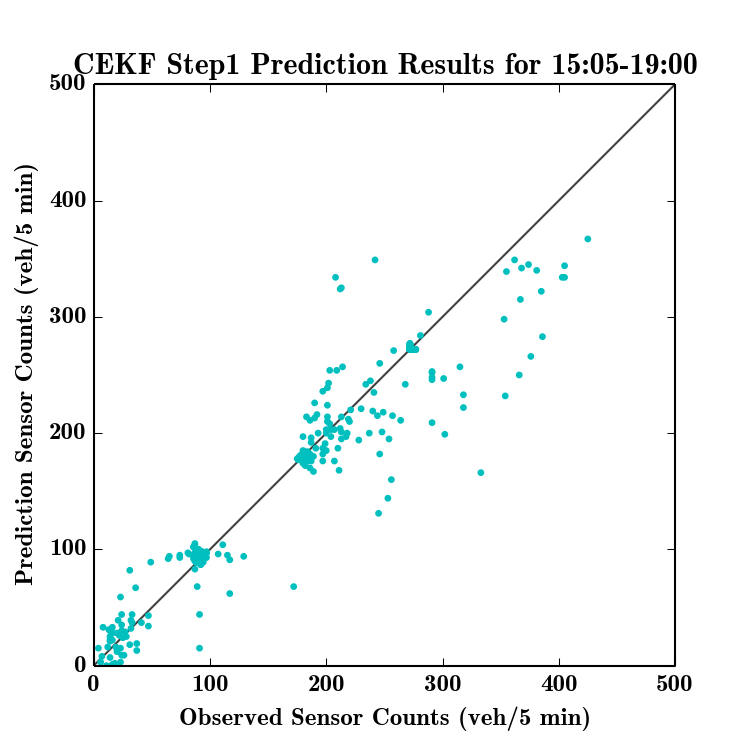}
			\includegraphics[width=\linewidth]{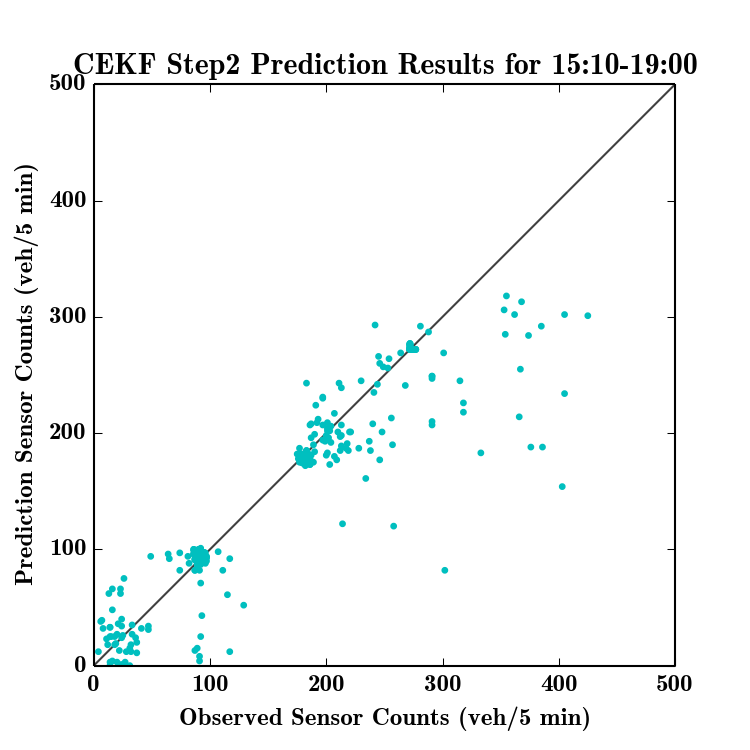}
			\includegraphics[width=\linewidth]{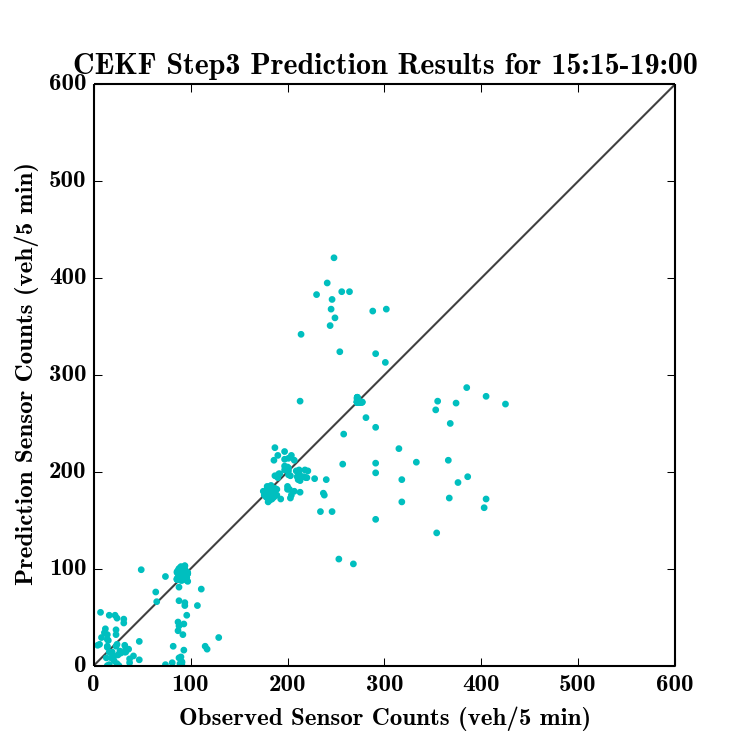}
		\end{minipage}%
		\begin{minipage}[t]{0.33\textwidth}
			\includegraphics[width=\textwidth]{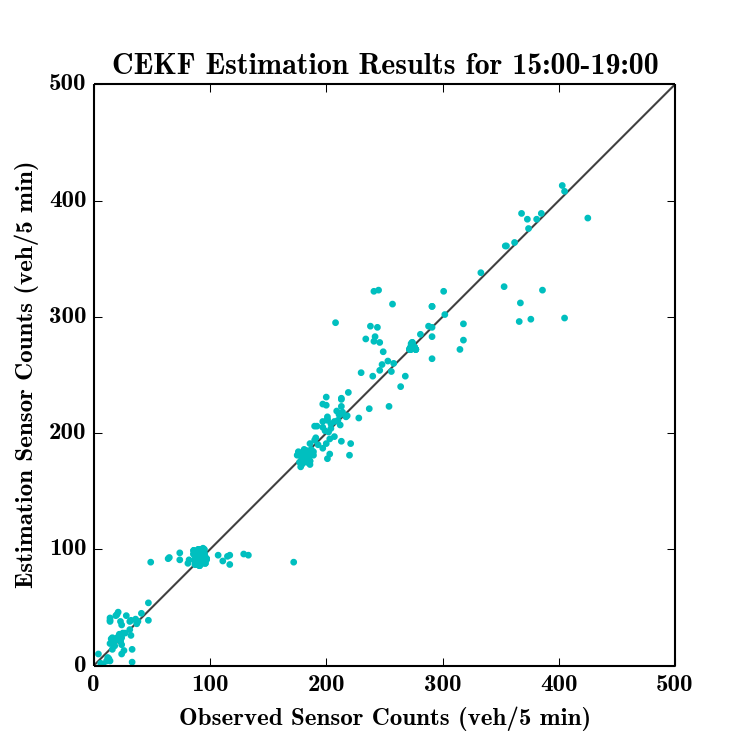}
			\includegraphics[width=\textwidth]{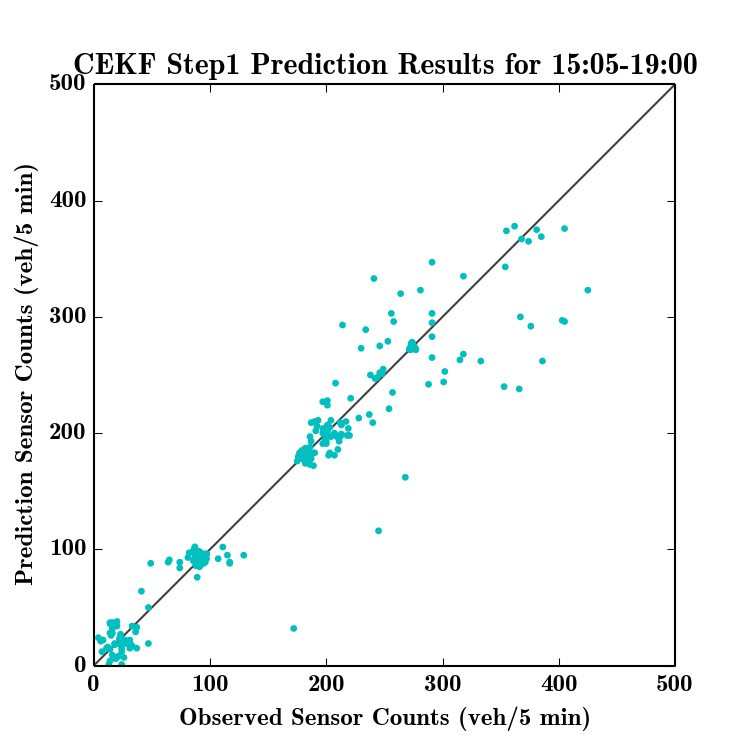}
			\includegraphics[width=\linewidth]{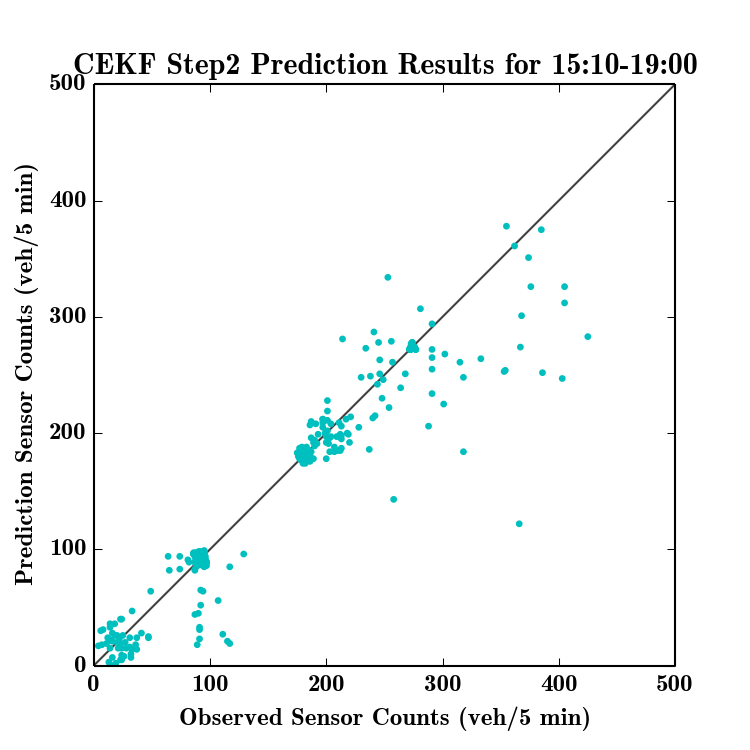}
			\includegraphics[width=\linewidth]{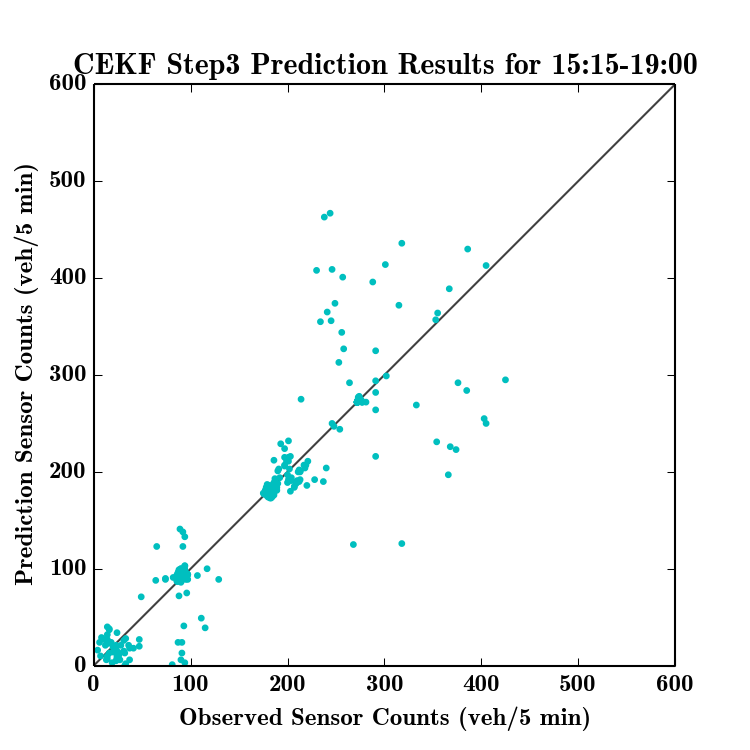}
		\end{minipage}%
	}%
	\caption{Scatter plot for estimated/predicted vs observed flow counts: left: CEKF(1), middle: CEKF(2), right: CEKF(5)}
	\label{fig:scatter}
\end{figure}

An in-depth performance comparison is given by \Cref{fig:scatter}. It presents the scatter plots of the estimated flows vs observations for each 5-minute interval during 15:00-19:00. Points closer to the diagonal line indicate a better fit. For the state estimation result in the first row, points in CEKF(2) (middle) and CEKF(5) (right) are closer to the diagonal line than CEKF(1), especially for counts less than 200 vehicles per interval. CEKF(5) also has slightly worse fit for observed counts greater than 350 vehicles per interval, when compared with CEKF(2) and CEKF(1). We think this may be a side effect of updating previously calibrated ODs to match delayed measurements; for CEKF(5), the states are updated 5 times to match the measurements in 5 intervals together. In other words, the augmented models trades off the accuracy of state estimations for predictions, which is discussed next.

The second to fourth rows show short-term predictions in the future 5, 10 and 15 minutes. CEKF(1) has more points below diagonal than the augmented models. This implies that CEKF(1) tends to underestimate the flow, because CEKF(1) is ``myopic'' and it only observes the OD flows' influence on measurements in the same interval. This situation is exacerbated during congestion. The estimated gradient is close to zero, because perturbing the input OD flows does not change the saturated flow rate. Thus, CEKF(1) is incapable of calibrating OD flows but to wait until the congestion dissipates near OD inputs.
In contrast, CEKF(5) captures the long-term effect of changing OD flows. Specifically, after perturbing OD flows, although the first-order gradient is zero (current interval), higher order gradients (prior intervals) still capture the impact of the perturbation resulting in a better fit in the prediction scatter plots. 

While we have shown the superior prediction power of state augmentation, we end this section with some comments on computational performance. The augmentation technique increases the dimension of matrices, making the EKF updates more cumbersome, and significantly increases the complexity of computing the gradient (Jacobian) as its size is a multiple of the nonaugmented gradient. The next section addresses how to speed up the computational performance.



\section{Partitioned Finite Difference Approach for Efficient Gradient Computation}\label{sec:PartSP}
We highlighted the importance of state augmentation that accounts for the delayed observation of hidden states, which is all-the-more relevant for congested networks. However, employing state augmentation will increase computational complexity and poses a critical challenge for real world applications (note that real-time performance is an issue even in case of the EKF with non-augmented states). 
For example, the online calibration procedure for each 5-minute interval in the case study in Section \ref{sec:CaseStudy} requires around 30 minutes of computational time, even without state augmentation on a 20-core server. A large proportion of the computational time involves gradient estimation (applying the finite difference method). Thus, the direct application of FD-EKF and state augmentation is impractical for real-time DTA systems on large networks. This section proposes an approach based on graph coloring to operationalize the partitioned finite difference method \citep{huang2010algorithmic} for improving computational performance.


Recall that each component of the gradient matrix $ \bs H_h^h $ in \Cref{eq:Grad_EKF} is $m$-by-$n$, corresponding to $m$ measurements and $n$ parameters (hidden states). These dimensions are the same across time intervals. The finite difference (FD) approach perturbs each parameter twice to obtain one column of the matrix $ \bs H_h^h $, thus requiring $2n$ runs of the simulator. The SPSA algorithm (reviewed in Section \ref{sec:LitRev}), which has been applied in the context of offline calibration, attempts to reduce the number of required function or simulator evaluations by performing a simultaneous perturbation of all the parameters, thus requiring only $2$ evaluations. However, inaccuracies in the gradient estimation arise from the fact that simultaneous perturbations of different parameters will have impacts on the same measurement. This will result in systematic overestimation or underestimation of the gradient. For instance, if perturbing two parameters cancels out their effects on a particular measurement, the associated gradients will be zero for both the parameters. Hence, it is preferable to perturb two parameters simultaneously only if they have no effect on any same sensor. The idea underlying this partitioned finite difference approach \citep{huang2010algorithmic} is to divide parameters into partitions such that in each partition, any two parameters do not simultaneously affect a measurement.

The partitioned finite difference approach we propose aims to approximate the gradient matrix with as few computations as possible, assuming knowledge of the gradient structure. It consists of three distinct steps: 
\begin{enumerate}
	\item \textbf{Gradient structure identification}: obtain an \textit{incidence} matrix $\boldsymbol{H}_{inc}$ to identify the sparse structure of the gradient matrix (Section \ref{sec:PSP_gradstruc}).
	\item \textbf{Parameter partitioning}: divide $n$ parameters into minimum $p$ partitions such that no two parameters in the same partition affect any common measurements. The parameters in these $p$ partitions should be mutually exclusive and collectively exhaustive (Section \ref{sec:PSP_part}).
	\item \textbf{Simultaneous perturbation for gradient estimation}: for each partition, perturb all parameters in two opposite directions to compute the gradient (Section \ref{sec:PSP_SP}).
\end{enumerate}

\subsection{Gradient Structure Identification}\label{sec:PSP_gradstruc}
The term \emph{structure} in this context refers to the locations of zeros and non-zeros in the gradient matrix. The gradient structure is necessary for the partitioning method to determine which parameters can be grouped together. An incidence matrix $\boldsymbol{H}_{inc}$ is a representation of the gradient structure. $\boldsymbol{H}_{inc,(i,j)}$ takes a value 1 if measurement $i$ and parameter $j$ are related and 0 if not. Formally,
\begin{align}
	\boldsymbol{H}_{inc,(i,j)} =\begin{cases}
		1 & \mbox{if }\boldsymbol{H}_{(i,j)}\neq 0\\
		0 & \mbox{if } \boldsymbol{H}_{(i,j)}= 0
	\end{cases} 
\end{align}

There are two general comments we would like to make about the gradient structure. First, the partitioning relies on the sparse nature of the gradient. More sparsity implies fewer shared measurements among parameters, thus resulting in fewer partitions and fewer finite difference evaluations. Second, the gradient structure may change across intervals, as traffic condition changes. Hence, when we assume a gradient structure beforehand, it must encompass all possible structures across intervals as the structure may change when traffic builds up. In other words, the overall $\boldsymbol{H}_{inc}$ should be the result of an element-wise \emph{or} operation of $\boldsymbol{H}_{inc,h}$ for all intervals $h$. In this case, we only need to partition once before calibration.

\subsection{Parameter Partitioning}\label{sec:PSP_part}

Given the gradient incidence matrix, we are ready to perform the partitioning. The partitioning problem involves grouping non-conflicting parameters, which we restate as a graph coloring problem. A heuristic solution procedure is then applied to solve the graph coloring problem. 

\subsubsection{Graph coloring problem}
Recall that the $j$th column of $\boldsymbol{H}_{inc}$ is the impact of the $j$th parameter on all the measurements. We want to group parameters that do not affect the same sensor. In other words, any two 1s in the same row disqualify grouping of the two corresponding parameters. The term \emph{conflict} is used to describe the fact that two parameters affect the same sensor measurement and we term these rows as \emph{conflicting} for a given pair of parameters. In a graphical representation, we denote the parameters as nodes, and each pair of nodes that has \emph{conflicting} rows are connected by edges. 

\begin{figure}[H]
	\begin{minipage}{0.6\textwidth}
		\centering
		\begin{align*}
			\boldsymbol{H}_{inc}=\begin{bmatrix}
				1 &1 &0 &0 &0 &1\\
				1 &1 &1 &0 &0 &0\\
				0 &1 &1 &1 &0 &0\\
				0 &0 &1 &1 &1 &0\\
				0 &0 &0 &1 &1 &1\\
				1 &0 &0 &0 &1 &1
			\end{bmatrix}
		\end{align*}
	\end{minipage}
	\begin{minipage}{0.4\textwidth}
		\includegraphics[width=.7\textwidth]{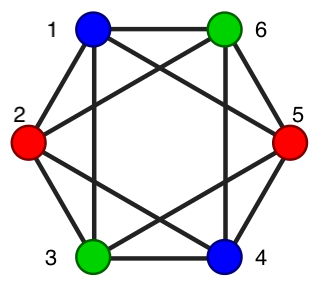}
	\end{minipage}
	\caption{A gradient incidence matrix and its corresponding optimal graph with 3 colors}
	\label{fig:graph_color}
\end{figure}

In this regard, the partitioning problem is equivalent to finding minimum colors for all the nodes such that no two connected nodes have the same color. For example, \Cref{fig:graph_color} presents a gradient incidence matrix, and the corresponding graph representation. The first row in $\boldsymbol{H}_{inc}$ shows that Node 1, 2 and 6 are connected, thus must be assigned with different colors. 

At this point, we would like to briefly discuss similarities and differences of this approach compared to variants of the SPSA algorithm. Despite the overt similarity, the variants of the SPSA including c-SPSA \citep{tympakianaki2015c}, w-SPSA \citep{lu2015enhanced} and PC-SPSA \citep{qurashi2019pc} do not in fact identify means of partitioning the parameters in the manner that we wish to do (i.e. ensuring that any two parameters in the same partition do not jointly affect a measurement). The c-SPSA performs clustering on the parameter values to identify clusters of homogenous parameters. Specifically, the authors focus on homogeneity of clusters defined in terms of the magnitude of OD flows (and not on their impacts on measurements), which helps address issues of scaling and allows for the definition of cluster specific gain sequences, the latter of which is specific to the SPSA. Similarly, the weight matrix in w-SPSA \citep{lu2015enhanced}, which is used to limit gradient noise, could potentially be used to identify partitions, but no method is proposed to do so by the authors. Likewise, the PC-SPSA does not provide a means of partitioning, but rather seeks to limit gradient noise and improve computational performance of the SPSA through dimensionality reduction. The application of the graph coloring method provides a systematic means of identifying a minimum number of partitions (the number of partitions does not have to be pre-specified), given knowledge of the gradient structure. By carefully choosing the perturbation partitions, our method would theoretically eliminate the impacts of the simultaneous perturbation, resulting in a gradient matrix without any loss in rank.

\subsubsection{Sequential/greedy graph coloring algorithm}
The problem of determining the minimum number of colors (or chromatic number) is known to be NP-hard \citep{coleman1983estimation}. Numerous heuristics have been proposed to determine the optimal coloring. We apply a sequential graph coloring algorithm from \citet{coleman1983estimation} that does not guarantee optimality, but has been widely used and analyzed.

\begin{algorithm}
	\begin{algorithmic}[htbp]
		\caption{Sequential Graph Coloring} \label{alg:GC}
		\State {
			$v_1, v_2, ..., v_n$ are $n$ nodes in the graph;\\
			$p=0$;}
		\For{$k$ = 1 to $n$}
		\For{$i$ = 1 to $p+1$}
		\If{connected nodes of $v_k$ not assigned Color $i$}
		\State{Break;}
		\EndIf
		\EndFor
		\State{Assign Color $i$ to $v_k$;}
		\If {$i>p$}
		\State{$p \gets p+1$;}
		\EndIf
		\EndFor
	\end{algorithmic}
\end{algorithm}

The resulting $p$ from the algorithm is the number of colors or partitions. This is a greedy algorithm, and the literature has reported that performance depends on the ordering of nodes (k loop in Algorithm \ref{alg:GC}). According to \citet{coleman1983estimation}, there exists an ordering of nodes such that the sequential graph coloring method can obtain the optimum. In our implementation, we perform the partitioning offline. Specifically, we run the sequential graph coloring algorithm with multiple random orderings of the nodes, and use the color assignment that attains the minimum. 

\subsubsection{Condensing the sparse gradient}
In order to formalize the gradient condensing process, assume the graph color assignments are in an $n$-by-$p$ zero-one matrix $\boldsymbol D$. The $j$th row indicates the color assignment of parameter $j$, where the $k$th element is 1 if $j$ is assigned to color $k$. The $k$th column $\boldsymbol{D}_k$ indicates all the parameters with color $k$. Since one parameter cannot be assigned to multiple colors, each row has exactly one element with value 1. 
The condensed gradient is given by:
\begin{align}
	\tilde{\boldsymbol{H}}_{(m\times p)} = \boldsymbol{H}\boldsymbol{D}
\end{align}


\subsubsection{Inflating the condensed gradient}
The sparse gradient $\boldsymbol{H}$ can be recovered without loss from condensed gradient $\tilde{\boldsymbol{H}}$ with the help of gradient incidence matrix $\boldsymbol{H}_{inc}$ as follows:
\begin{align}
	\boldsymbol{H}_{(m\times n)}= \left(\tilde{\boldsymbol{H}}\boldsymbol{D}^\top\right) \circ \boldsymbol{H}_{inc} 
	\label{eq:inflate_gradient}
\end{align}
where, $\circ$ is the element-wise product. 

\subsection{Simultaneous Perturbation for Gradient Estimation}\label{sec:PSP_SP}
Given the partitioning of the parameter vector, the $k$th column of the condensed gradient $\tilde{\boldsymbol{H}}$ can be computed using simultaneous perturbations as:
\begin{align}
	\tilde{\boldsymbol H}_k &= \frac{\boldsymbol g_h(\hat{\boldsymbol{x}}_{h|h-1}+\boldsymbol\delta_k) - \boldsymbol g_h(\hat{\boldsymbol{x}}_{h|h-1}-\boldsymbol\delta_k)}{2\delta_k}  \\
	\boldsymbol \delta_k &= \delta_k \boldsymbol{D}_k \mbox{  \ \ \ \ } \forall k=1,2,...,p\\
	\boldsymbol{H} &= \left(\tilde{\boldsymbol{H}}\boldsymbol{D}^\top\right) \circ \boldsymbol{H}_{inc} 
\end{align}
where, $\delta_k$ is the perturbation size for all parameters in the same partition, $\boldsymbol{D}_k$ is the $k$th column of matrix $\boldsymbol{D}$. 

%
%
\subsection{Factors Affecting the Performance of Partitioning}

As mentioned above, the sequential/greedy method is not guaranteed to reach optimality. We partly address this issue by using the best result from multiple runs. In this section, we discuss some factors that determine the optimal chromatic number for the OD partitioning problem. 

First, the problem clearly has dependencies on sensor placement within the network. We consider an example for the OD calibration problem, and assume all the OD pairs are captured by flow count sensors already. Adding more sensors to the same region would likely increase the number of partitions needed because there will be more \emph{conflicting} rows. This is equivalent to adding more rows in the incidence matrix and more links between nodes in \Cref{fig:graph_color}. Furthermore, the partitioning process is closely related to the problems of sensor location and observability (see \cite{castillo2015state,castillo2008observability,
	yang2018stochastic,ehlert2006optimisation})

Second, it is trivial but crucial to notice that there are trips that do not share any segments, in which case we can safely assign them to the same partition. This is reflected in the sparse nature of the gradient matrix, which has been observed and exploited for DTA calibration in other contexts \citep{wen2008scalability}. Thus, assuming sensors are uniformly distributed across the network, regardless of their density, one can reasonably expect there are gains from the partitioning since it likely that they will always be sensors unaffected by certain OD flows due to the typical spatial patterns of demand and sensors in urban areas.

Finally yet importantly, the partitioning result will be affected by the degree of augmentation $r$. Since we use staggered horizons (as seen in \Cref{ssec:staggered_horizons}), we need to make sure the parameters in a partition do not have \emph{conflicts} for all $r$ gradients that are obtained within one simulation run. Similarly, the H matrix in \Cref{eq:inflate_gradient} will be vertically concatenated and have dimension $rm$ by $n$. Thus increasing $r$ would also increase the chromatic number and decrease the effectiveness of the partitioning.

\subsection{Performance on a Real Network}
In this section, we conduct an experiment to demonstrate the performance of PSP-EKF (EKF with the gradient computed using the partitioned simultaneous perturbation approach) on the Singapore Expressway network displayed in \Cref{fig:sgpmap}. For more details on the network, refer Section \ref{sec:CS_Data}. 
%

We once again consider the online OD estimation problem (where parameters are dynamic OD demands and measurements are sensor flow counts). Real traffic counts on a selected day in 2015 (provided by the Land Transport Authority) are used, the historical OD demands are obtained from a prior offline calibration. An assumed AR process is used for the transition equation.  In order to obtain a universal gradient incidence matrix throughout the whole simulation period, we first perform online calibration using the FD-EKF algorithm and record all the H matrices. As mentioned before, the incidence matrix is computed with element-wise \emph{or} on all the incidence matrices from each interval. The sequential graph coloring algorithm (Section \ref{sec:PSP_part}) is applied and generates 438 partitions. 

The accuracy and computational performance of the PSP-EKF are now compared for a simulation period for 7-10AM, with 5-minute OD intervals. The supply parameters are fixed across the two algorithms and the simulations are run on a server with 40 cores. The RMSN in sensor counts across all intervals is shown in \Cref{tbl:pspekf_comp} and indicates that the performance for both state estimation and prediction is similar for both methods.
\begin{table}[hbt!]
	\caption{Calibration accuracy comparison for FD-CEKF and PSP-CEKF }
	\centering
	\begin{tabular}{ccccc}
		\Xhline{1pt}
		\multirow{2}{*}{Method} & \multirow{2}{*}{\begin{tabular}[c]{@{}c@{}}Estimation\\ RMSN\end{tabular}} & \multicolumn{3}{c}{Prediction RMSN} \\
		&     & 1 step     & 2 step     & 3 step    \\ \hline
		No calibration          & 59.7\%    & 59.7\%      & 59.7\%      & 59.7\%     \\
		FD-CEKF                 & 32.1\%    & 34.0\%      & 36.3\%      & 38.3\%     \\
		PSP-CEKF                & 32.9\%    & 34.7\%      & 37.0\%      & 39.0\%    \\ \Xhline{1pt}
	\end{tabular}
	\label{tbl:pspekf_comp}
\end{table}

In terms of computational performance, the traditional central finite difference (FD-EKF) requires 4121 pairs of simulations to estimate the gradient in each interval  compared to 438 for the PSP-EKF. This is reflected in the significant computational gains (on average, a six fold decrease in computational time) shown in \Cref{tbl:psp_comp}. The gradient structure detection needs a full run of the FD-CEKF so the time is similar to the first row in the table. The graph-coloring algorithm then uses the gradients and finishes quickly (within 5 minutes for 30 runs with random initialization).

\begin{table}[hbt!]
	\caption{Computation time comparison for FD-CEKF and PSP-CEKF iterations}
	\centering
	\scriptsize
	\begin{tabular}{cccccc}
		\begin{tabular}{ccccccc}
			\Xhline{1pt}
			\multirow{2}{*}{\begin{tabular}[c]{@{}c@{}}Estimation\\ method\end{tabular}} & \multirow{2}{*}{\begin{tabular}[c]{@{}c@{}}\# Parameter\\ groups\end{tabular}} & \multicolumn{5}{c}{Calibration time for interval (minutes)} \\
			& & 6:00-6:05  & 7:00-7:05  & 8:00-8:05  & 9:00-9:05  & Average \\ \hline
			FD & 4121 & 12.2       & 22.3       & 32.6       & 48.3       & 28.8    \\ \hline
			PSP & 438  &  2.5  & 3.9 & 5.3 & 7.2 & 4.7 \\ \Xhline{1pt}
		\end{tabular}
	\end{tabular}
	\label{tbl:psp_comp}
\end{table}

In summary, the PSP-EKF approach attains a very similar accuracy as the FD-EKF whilst significantly improving computational performance (real-time for a five minute estimation interval) with the extent of improvement depending on the sparsity of the gradient structure. This will allow the online calibration to be real-time operational with a moderate amount of parallelization.

\section{Case Study}\label{sec:CaseStudy}
In this section we conduct experiments on a large-scale network for the online OD estimation problem using real world data to examine the performance of the approaches proposed in this paper; in particular, state augmentation and partitioned simultaneous perturbation. Section \ref{sec:CS_Data} introduces the experimental setting and Section \ref{sec:CS_results} discusses the results and findings.

\subsection{Experimental Setup}\label{sec:CS_Data}

\begin{figure}[H]
	\centering
	\includegraphics[width=0.9\linewidth]{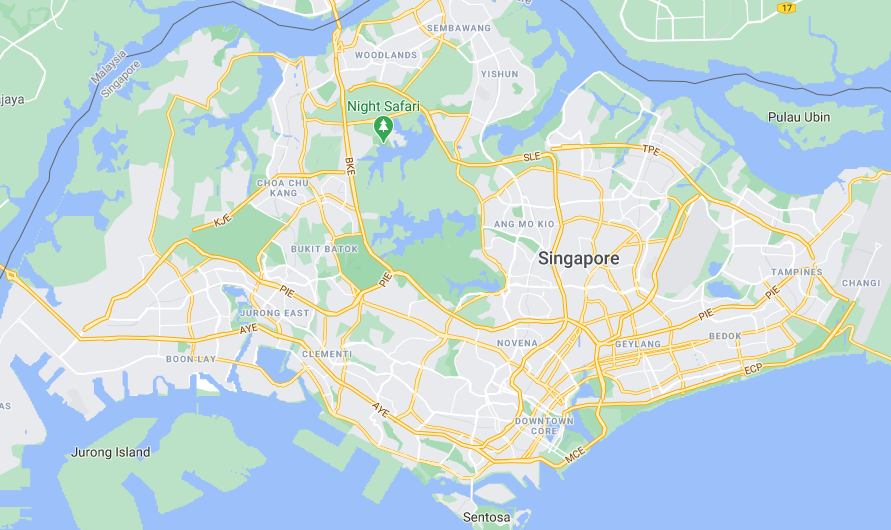}
	\caption{Singapore expressway network \citep{googlemap2020sgp}}
	\label{fig:sgpmap}
\end{figure}

\begin{figure}[H]
	\centering
	\includegraphics[width=0.8\linewidth]{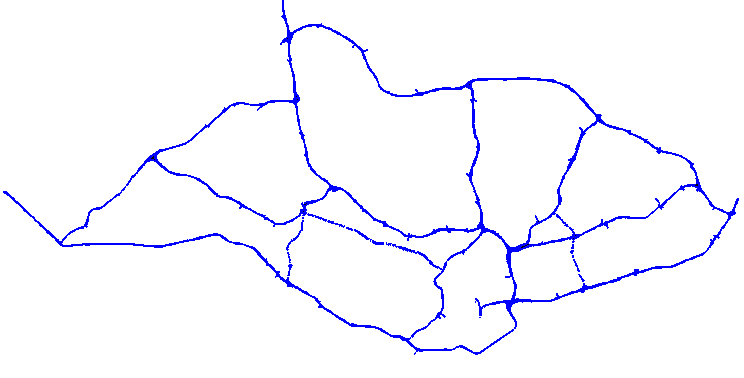}
	\caption{Singapore expressway network in the DTA model}
	\label{fig:sgpnetbw}
\end{figure}

The Singapore expressway network is a large-scale city-wide urban network shown in \Cref{fig:sgpmap}. The corresponding representation of the network used in DynaMIT is shown in \Cref{fig:sgpnetbw}. It includes all the expressways and selected arterials. The network consists of 939 nodes, 1157 links and 3906 segments. There are 4121 origin destination (OD) pairs on the network (OD flows are discretized in five minute intervals), where on-ramps serve as origin nodes and off-ramps serve as destination nodes. These 4121 OD pairs have 18532 routes in total and further, there are 650 sensors (measurements at five minute intervals) distributed across the network that capture traffic flow volumes. 

Real-time traffic flow volumes (counts) on the 650 sensors are provided by the Land Transport Authority (LTA) in Singapore for 14 weekdays in December 2015.
In this case study, the OD estimation problem is considered once again for the morning peak period from 6AM to 10AM. The main objective is to examine the performance of the augmented state space model using real-time traffic flow measurements. We also applied partitioned simultaneous perturbation for all the experiments to speed up gradient computations.  

The calibration parameters are 4121 OD demands for each 5-minute departure interval. Route choice and supply parameters such as speed-density relationships and capacity for each segment are set to offline calibrated values. Before conducting the experiments, the Kalman filtering framework needs several additional inputs, namely:

\begin{itemize}
	\item Time-dependent Historical OD matrices
	\item The autoregressive (AR) model for the transition equation 
	\item The transition and measurement error covariances $\bs Q$ and $\bs R$
\end{itemize}

\begin{figure}[H]
	\centering
	\includegraphics[width=0.8\linewidth]{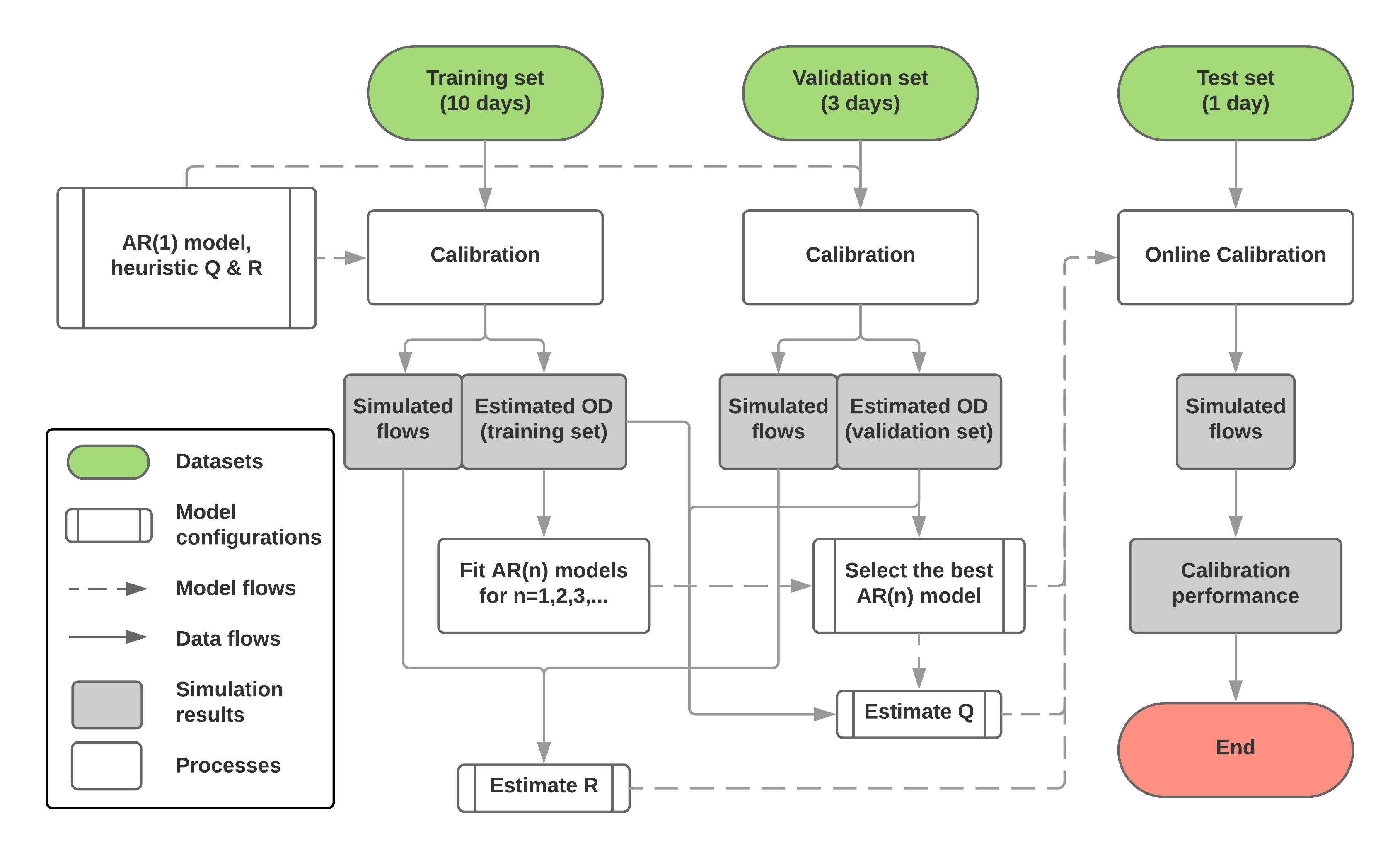}
	\caption{Framework for computational experiments}
	\label{fig:Test_Process}
\end{figure}

The flowchart in \Cref{fig:Test_Process} summarizes the procedure to obtain these inputs and is described next.

\begin{enumerate}
	\item Divide the weekdays into training set (10 days), validation set (3 days) and test set (1 day);
	\item Perform calibration using the FD-CEKF (constrained extended Kalman filter) algorithm for the training and validation set. CEKF is applied to properly model the non-negativity constraint for OD flows \citep{zhang2017improved}. The covariance matrices and transition equation used in this stage of calibration are based on heuristics. Specifically, the values of $\bs Q$ were initially set by assuming a diagonal structure and a coefficient of variation of 10\% (mean values were obtained from a seed OD, i.e. an OD matrix from a prior calibration). Along similar lines, $\bs R$ was set by assuming a diagonal structure and a coefficient of variation of 10\% (mean values were observed historical sensor counts).
	
	\item Calculate the residuals between the estimated flows and the observed data in the training set and validation set. We then compute the variance of the residuals for each sensor across time intervals. These variances serve as diagonal elements of $\bs R$. The calculation is given by:
	\begin{align}
		\bs R &= \begin{bmatrix}
			r_1 &0 &\cdots &0\\
			0 &r_2 &\cdots &0\\
			\vdots &\vdots &\ddots &\vdots\\
			0 &0 &\cdots &r_m
		\end{bmatrix} \label{eq:R}\\
		r_i &= \frac{1}{DN} \sum\limits_{d=1}^{D} \sum\limits_{h=1}^{N} \left(M_{h,i}^{(d)} - g_{h,i}^{(d)}(\cdot) \right)^2 \label{eq:r}
	\end{align}
	where, $M_{h,i}^{(d)}$ is the observed measurement and $g_{h,i}^{(d)}(\cdot)$ is the simulated flow estimate for the $i$th sensor at time interval $h$ on day $d$. In total we have $D$ days of training and validation data.
	
	\item Fit an AR(\textit{n}) model to calibrated time-dependent OD matrices in the training set. For each \textit{n} taking a value from 1 to 5, we fit an AR model. Then we test the models and select the \emph{best} model based on their prediction performance on the validation set. The best model from the training and validation sets is an AR(2) model, given by:
	\begin{align}
		\bs x_h = 0.884\bs x_{h-1} + 0.0967\bs x_{h-2} + \epsilon \\
		\epsilon \sim \mathcal{N}(\bs 0, 17.6 \bs I)
	\end{align}
	
	The variance magnitude $q = 17.6$ was obtained from the following procedure. First, we calculate the residuals between each estimated OD and predicted values given by the model above. Then we compute the variance of the residuals for all ODs across time intervals. It serves as a universal variance for all the diagonal elements of $\bs Q$:
	\begin{align}
		\bs Q =& \begin{bmatrix}
			q &0 &\cdots &0\\
			0 &q &\cdots &0\\
			\vdots &\vdots &\ddots &\vdots\\
			0 &0 &\cdots &q
		\end{bmatrix} \label{eq:Q}\\
		q = \frac{1}{DnN} \sum\limits_{d=1}^{D}& \sum\limits_{i=1}^{n}\sum\limits_{h=1}^{N} \left(\left(\bs X_{h}^{(d)}\right)_i - \left(\bs\Phi\bs X_{h-1}^{(d)}\right)_i \right)^2 \label{eq:q}
	\end{align}
	where, $\bs X_{h}^{(d)}$ is the state vector in interval $h$ on day $d$. $\left(\bs X_{h}^{(d)}\right)_i$ is the $i$th calibrated OD, and $\left(\bs\Phi\bs X_{h-1}^{(d)}\right)_i$ is the $i$th predicted OD with the AR model, parameterized by $\bs\Phi$. $n$ is the number of OD pairs.
	\item The computed $\bs Q$ and $\bs R$, together with the selected AR(2) model serve as inputs for the Kalman filter in online calibration. 
	\item The mean of the calibrated demand over the training and validation set for each interval serves as the time-dependent historical values to construct deviations for the test set.
\end{enumerate}

\begin{figure}[H]
	\centering
	\includegraphics[width=.55\linewidth]{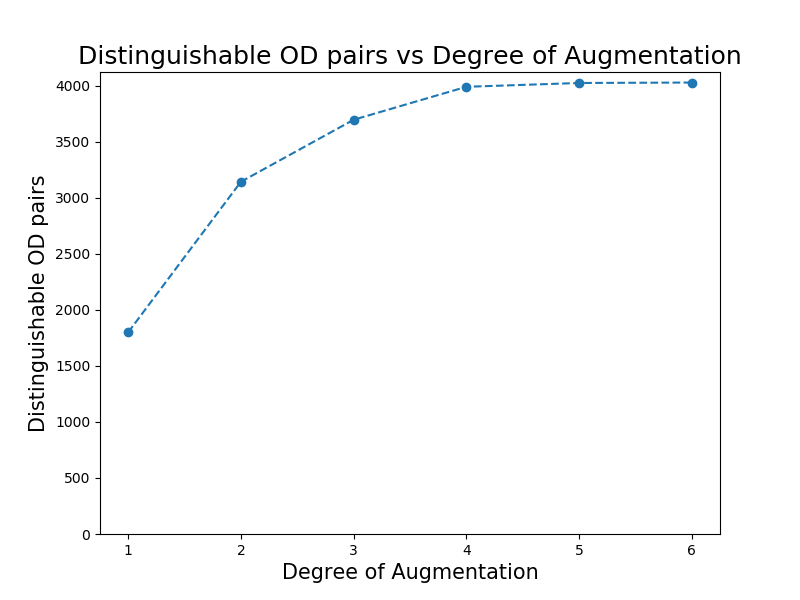}
	\caption{Distinguishable OD pairs versus state augmentation degree}
	\label{fig:DisOD}
\end{figure}

The experiments examine the performance of the original state space model and the augmented model of various degrees. We applied parameter partitioning for all the augmented models to accelerate the calibration process. We also apply CEKF to model the non-negative OD flows. We consider the following configurations. 
\begin{enumerate}[]
	\item \textbf{CEKF(1)}: constrained extended Kalman filter without state augmentation;
	\item \textbf{CEKF(3)}: constrained extended Kalman filter with state augmented to degree 3;
	\item \textbf{CEKF(6)}: constrained extended Kalman filter with state augmented to degree 6.
\end{enumerate}
In addition, we consider a benchmark that involves directly using the historical demand (from the training and validation sets) without online calibration. The choice of the augmentation degrees of 3 and 6 are based on a heuristic that computes an approximate measure of observability (note that this is not a rigorous definition of observability---see \cite{castillo2015state,castillo2008observability,
	yang2018stochastic}---but rather ensures that there are no hidden states due to delayed measurements) as a function of augmentation degree. For a given degree of augmentation, we try to identify the number of OD pairs that are distinguishable assuming we only observe their impacts on the measurements up to that degree (i.e. within the augmented intervals). This is approximated using knowledge of network link travel times, the shortest paths between each OD pair, and the location of sensors. Specifically, for a degree of augmentation $r$, an OD pair is termed \textit{distinguishable} if the first sensor at which it can be identified (amongst OD flows originating from the same node) is reachable from the origin within $r$ intervals. This is shown in Figure \ref{fig:DisOD} where a degree of 3 and higher are sufficient to minimize the impacts of hidden states due to delayed measurements.

\subsection{Results and Discussion}\label{sec:CS_results}
In this section, we discuss the calibration accuracy of each model quantified with three measures: the root mean square error (RMSE), weighted sum of squared error (WSSE) and root mean squared normalized error (RMSN). The lower values of these three metrics indicate a better accuracy of the experiment. As a metric to address the sensors that have a high measurement error, the WSSE utilizes the inverse of $\bs R$ as weights for the squared errors of each measurement. In our case of a diagonal $\bs R$, each squared error is divided by its assumed variance and then summed up. Thus, the WSSE discounts the impact of the uncertain measurements. Also, note that the WSSE is part of the Kalman filter's objective function \citep{sorenson1970least}, and thus a lower bound of it. 

\subsubsection{Calibration Accuracy and Prediction Performance}
\Cref{tbl:rmsn_all} summarizes the accuracy of each approach for state estimation and prediction. First, all CEKF configurations significantly improve over the historical benchmark. With regard to state estimation, CEKF(3) obtains the lowest error in terms of RMSN, with relative improvements of 13\% with respect to CEKF(1). The CEKF(6) on the other hand yields a higher RMSN for estimation than CEKF(3), but still shows a 4\% relative improvement over CEKF(1). However, in terms of the WSSE, the augmented models perform worse than the non-augmented model. This is discussed later in the section.  

\begin{table}[H]
	\centering
	\caption{Performance of all experiments on test day, simulation period 6:20-7:20}
	\small
	\begin{tabular}{cccccccc}
		\Xhline{1pt}
		\multirow{2}{*}{Index} & \multirow{2}{*}{Description} & \multicolumn{3}{c}{Estimation} & \multicolumn{3}{c}{Prediction RMSN} \\
		&                              & RMSE   & WSSE     & RMSN       & 1 step     & 2 step     & 3 step    \\ \hline
		0          & Historical   & 112.6   & 18047  & 36.6\%     & 36.3\%     & 36.2\%     & 35.9\%    \\
		1         & CEKF(1)        &  109.7   & 13664 & 33.1\%     & 33.9\% &  34.9\%     &  34.4\%    \\
		2         & CEKF(3)        & 106.8  & 13995  & 28.7\%     & 30.1\%     & 31.1\%     & 30.1\%    \\
		3         & CEKF(6)        & 111.0  & 16409   & 31.7\%     & 30.7\%     & 31.9\%      & 30.8\%         \\ \hline
		\multirow{2}{*}{Index} & \multirow{2}{*}{Description} & \multicolumn{3}{c}{Prediction RMSE} & \multicolumn{3}{c}{Prediction WSSE} \\ 
		&       & 1 step     & 2 step     & 3 step   & 1 step     & 2 step     & 3 step    \\ \hline
		0          & Historical   & 116.4   & 120.6  & 124.1     & 19064     & 20022     & 20924    \\
		1a         & CEKF(1)         &  114.6   & 119.8 & 123.2    & 17428 &  19133     &  21013    \\
		2a         & CEKF(3)        & 109.7  & 115.0  & 118.2     & 16498     & 18165     & 20007    \\
		3a         & CEKF(6)        & 110.3  & 116.3   & 119.1    & 16969    & 18632      & 20716         \\  \Xhline{1pt}
	\end{tabular}
	\label{tbl:rmsn_all}
\end{table}

In terms of state prediction performance, the results in \Cref{tbl:rmsn_all} indicate an improvement of around 13\% in RMSN for the augmented models compared with CEKF(1), with CEKF(3) again yielding the best performance. The prediction errors in terms of RMSE and WSSE also improve for the augmented models relative to the non-augmented CEKF(1). Moreover, the non-augmented model shows a higher error in 3-step predictions (WSSE) compared to even the historical benchmark. Thus, state augmentation shows a clear improvement in prediction performance.

\begin{table}[H]
	\caption{Sensor RMSE grouped by variances in $\bs R$ for estimation and 3 step prediction}
	\label{tbl:var_grp}
	\centering
	\begin{tabular}{cccccccc}
		\Xhline{1pt}
		\multirow{2}{*}{$r$} & range from      & 1 & 500 & 2500 & 5000 & 10000 & 20000 \\
		& range to  & 500 & 2500 & 5000 & 10000 & 20000 & $+\infty$ \\ \hline
		\multicolumn{2}{c}{Number of sensors}                                                  & 118            & 111            & 90             & 115            & 109             & 107            \\ \hline
		\multirow{3}{*}{Estimation}                                                  & CEKF(1) & 35.48          & 53.86          & \textbf{76.72} & \textbf{132.1} & 125.1           & 157.2          \\
		& CEKF(3) & \textbf{35.28} & \textbf{52.67} & 81.99          & 134.6          & \textbf{119.2}  & \textbf{146.7} \\
		& CEKF(6) & 38.87          & 56.49          & 88.60          & 140.4          & 123.6           & 149.5          \\ \hline
		\multirow{3}{*}{\begin{tabular}[c]{@{}c@{}}3 step\\ prediction\end{tabular}} & CEKF(1) & 48.32          & 60.76          & 86.42          & 148.4          & 142.9           & 170.0          \\
		& CEKF(3) & \textbf{47.36} & \textbf{58.11} & 86.67          & \textbf{143.0} & \textbf{134.4}  & \textbf{163.0} \\
		& CEKF(6) & 48.59          & 58.39          & \textbf{86.20} & 143.8          & 136.1           & 165.5         \\ \Xhline{1pt}
	\end{tabular}
\end{table}

To further investigate the estimation performance of models in terms of WSSE, we divide the sensors into groups based on their assumed variances in $\bs R$ and examine the RMSE for each group (\Cref{tbl:var_grp}). The best RMSE for each group is shown in bold. We have two main observations. First, the CEKF(3) and CEKF(6) have similar prediction performances. The reason may lie in the fact that there are 4121 OD pairs and 650 sensors implying that a large degree of freedom exists even in the non-augmented model. While augmenting the states further increases the model complexity, the benefit may be marginal when the degrees of freedom are already large. The marginal improvement also implies a degree of 3 for augmentation is sufficient for this network. The second observation is that the major improvement with the augmented models lies in sensors with large assumed variances. Recall that these were estimated from residuals of non-augmented models on the training set. Thus, this observation indicates that augmented models may improve the sensors that were poorly fitted in non-augmented models. These improvements are clear and significant. 


\begin{figure}[H]
	\centering
	\includegraphics[width=.475\linewidth]{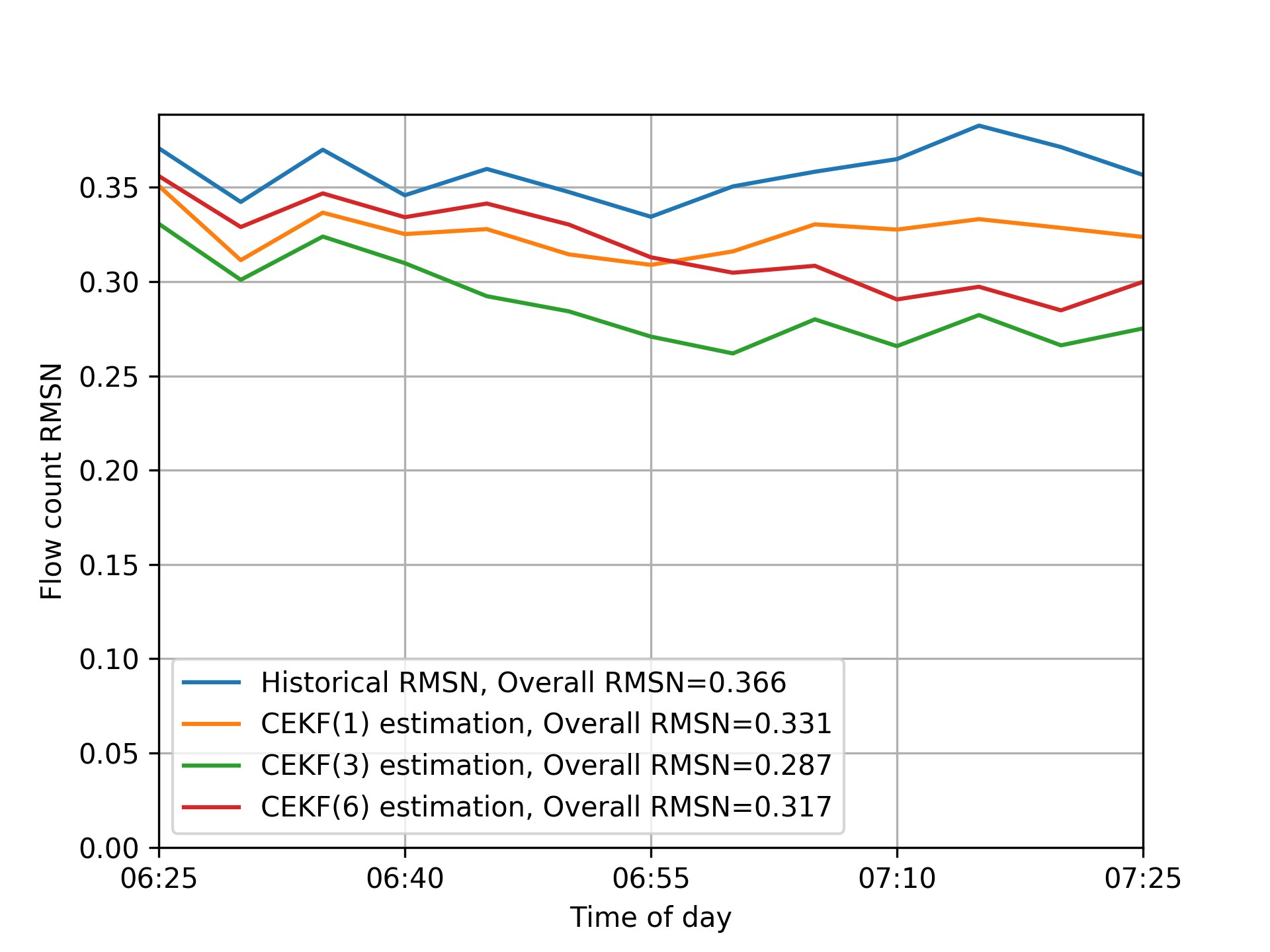}
	\includegraphics[width=.475\linewidth]{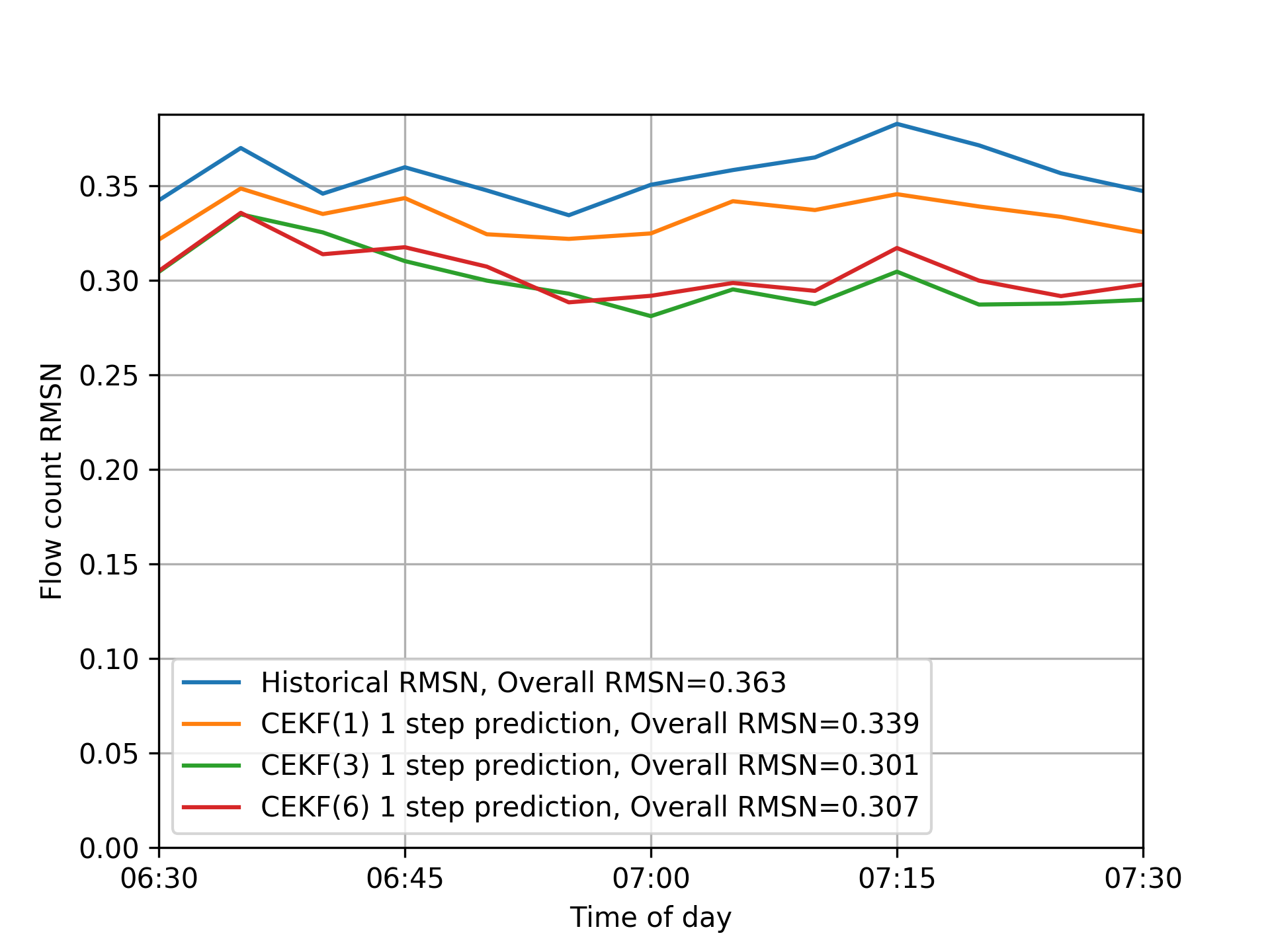}\\
	\includegraphics[width=.475\linewidth]{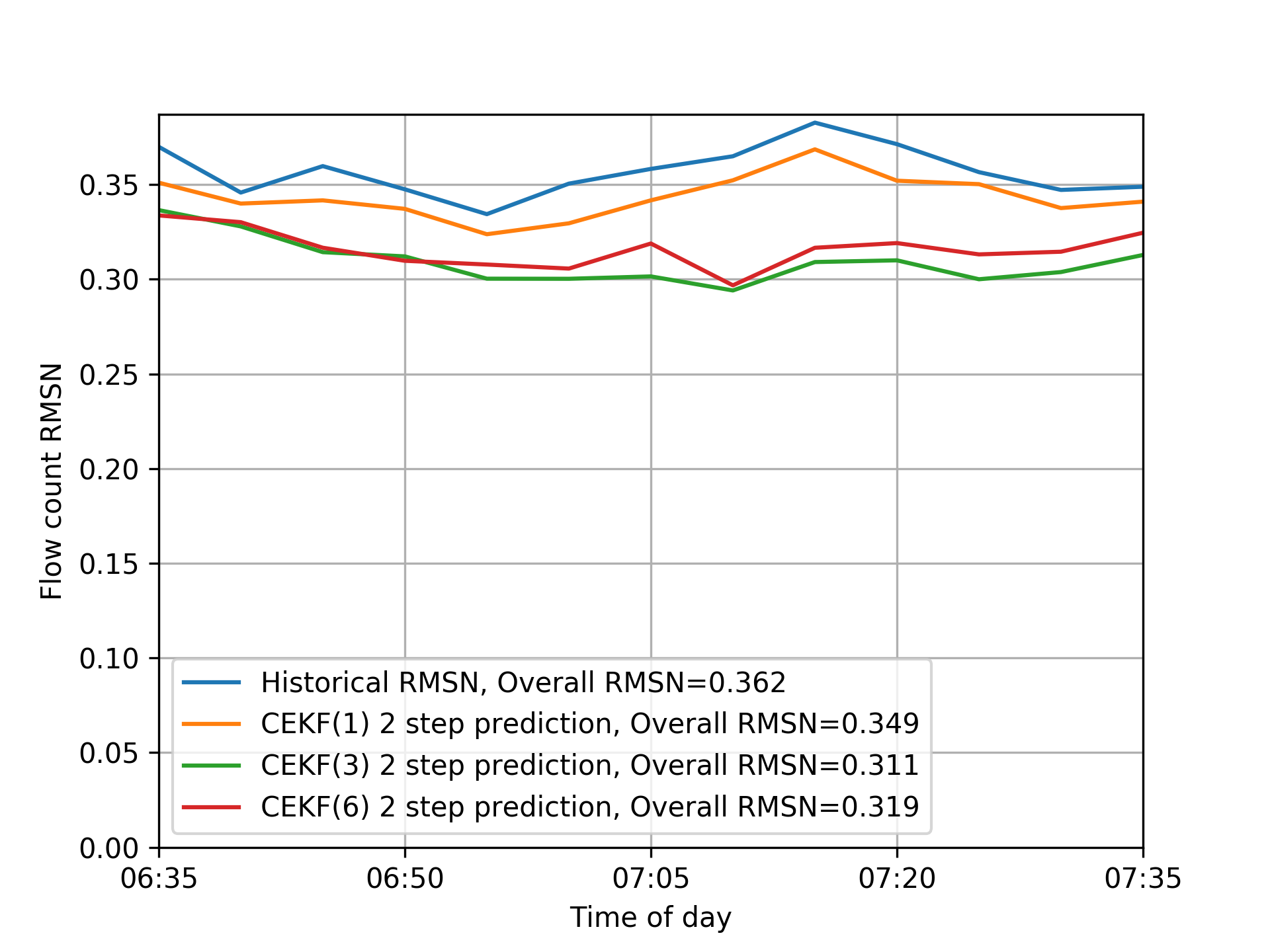}
	\includegraphics[width=.475\linewidth]{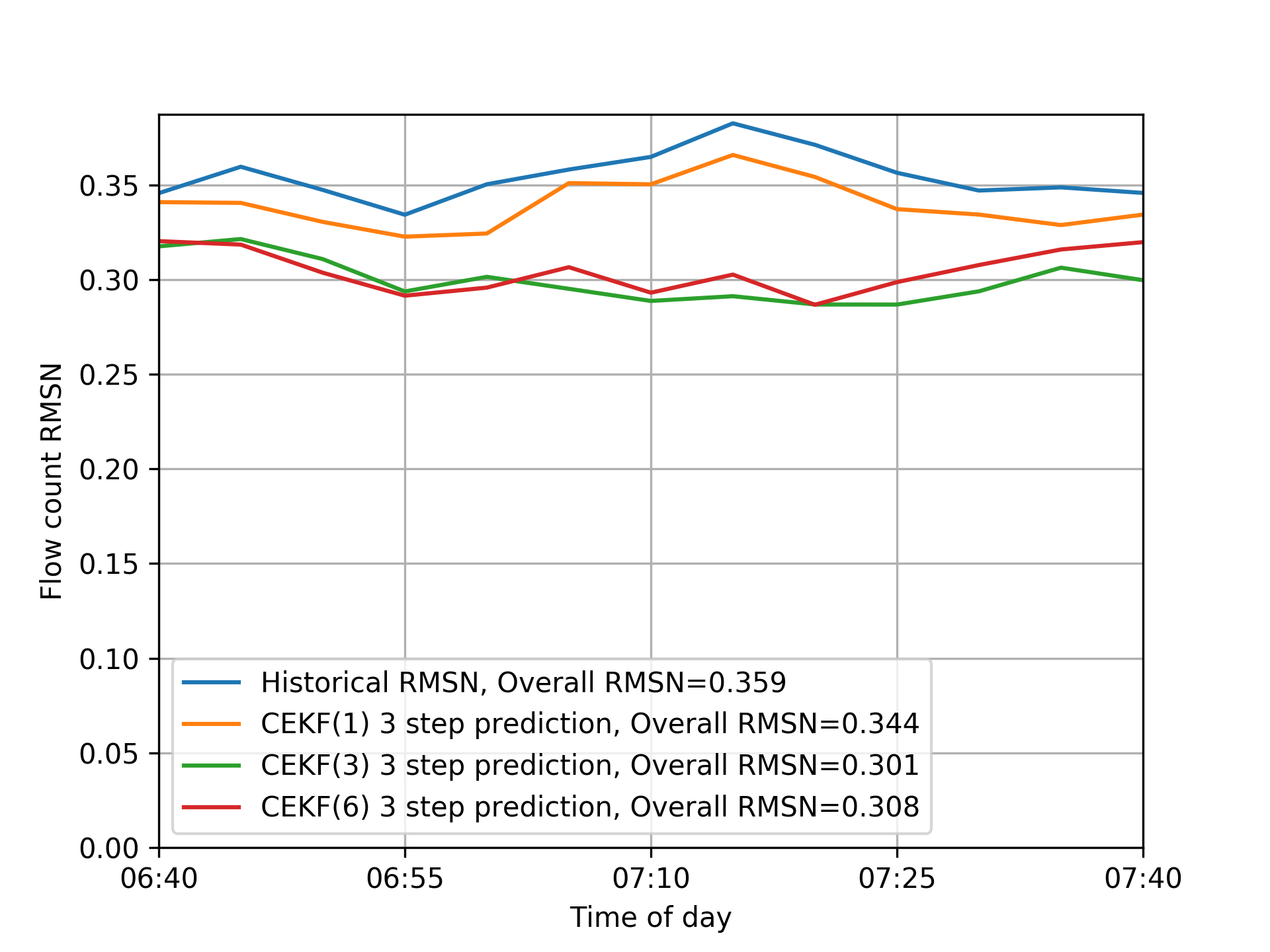}
	\caption{Flow volume RMSN for estimation (top left) and predictions, simulation period 6:20-8:20}
	\label{fig:RMSNcomparison}
\end{figure}

The time-dependent RMSN values for both estimation and prediction are shown in \Cref{fig:RMSNcomparison}. It is noticeable that augmented models give the better predictions than CEKF(1) by a significant and clear margin. 
Based on the discussion in Section \ref{sec:StateAug}, a natural explanation of the CEKF(1)'s performance is the inability to model the delay in measurements. The non-augmented model ignores the correlation between parameters and measurements across intervals and thus, previously estimated ODs cannot be adjusted. In other words, 
the model is forced to explain measurements with parameters in the same interval and the error term absorbs the effect of the omitted variables resulting in a less accurate model.  
In contrast, when we use an augmented model, part of the error in measurements now can be ``explained'' by modeling the effect of parameters in previous intervals. 
As a result, longer trips are captured in later intervals and hence can be estimated better with augmented models. Therefore, 
with a good AR model, predicted parameters will be more accurate, which yield better traffic predictions as the experiments corroborate. 

With regard to prediction accuracy, the augmented models significantly improve over CEKF(1). However, we also notice the improvement is slightly less when compared with the synthetic example in Section \ref{sec:StateAug}. This may be attributed to two reasons. First, as mentioned before, there are significantly more parameters (4121) than observations (650) in each interval. Given the large degrees of freedom in the problem, non-augmented models may perform well enough in terms of goodness-of-fit measures for sensors. Second, excessive noise in the real flow measurements leads to violations in flow conservation in some cases which has an effect on model performance. 

\subsubsection{Computational Performance}
In this section, we compare the computational performance of CEKF(1) (non-augmented model) with CEKF(3) for this case study. All experiments were conducted on an AMD Ryzen 9 3900X desktop computer using 20 cores (3600MHz DDR4 dual channel RAM).

\begin{table}[hbt!]
	\caption{Computational Time Comparison for CEKF(1) vs CEKF(3)}
	\label{tbl:complexity}
	\centering
	\begin{tabular}{c|cc|cc}
		\Xhline{1pt}
		\textbf{\begin{tabular}[c]{@{}c@{}}Traffic in\\ Network\end{tabular}} & \multicolumn{2}{c|}{\textbf{\begin{tabular}[c]{@{}c@{}}Total Computational\\ Time (minutes)\end{tabular}}} & \multicolumn{2}{c}{\textbf{\begin{tabular}[c]{@{}c@{}}CEKF Optimization\\ Time (seconds)\end{tabular}}} \\
		& CEKF(1)                                           & CEKF(3)                                          & CEKF(1)                                            & CEKF(3)                                            \\ \hline
		13000                                                                 & 2.4                                               & 11.6                                             & 1.9                                                & 70.1                                               \\
		25000                                                                 & 3.0                                               & 14.6                                             & 1.9                                                & 69                                                 \\
		41500                                                                 & 3.7                                               & 19.4                                             & 1.8                                                & 71                                                 \\
		55000                                                                 & 4.3                                               & 22.4                                             & 1.9                                                & 73                                                 \\
		73000                                                                 & 5.1                                               & 26.3                                             & 2.1                                                & 73                                                 \\ \Xhline{1pt}
	\end{tabular}
	
\end{table}

\Cref{tbl:complexity} compares the computational performance for two methods in each 5-minute interval when they are loaded with the same traffic in the network. The  CEKF(6) costs significantly higher computational time than CEKF(3), but yields insignificant improvements, hence the results of CEKF(6) are omitted. The cost increase is roughly in line with the analysis in Section \ref{ssec:staggered_horizons} and \ref{ssec:cekf_cost}. In these experiments, the results indicate that the CEKF(1) achieves real-time performance with computational times less than the typically used 5 minute inteval for state estimation. This is in large part a consequence of the roughly six fold improvement in computational times achieved due to the partitioned finite difference (PSP-EKF) approach presented previously. Although the CEKF(3) does not yield real-time performance (computational times of 11 minutes to 25 minutes) with the computational architecture employed in these experiments, a modest increase in the number of cores will allow for real time performance given that the gradient estimation---the computationally most intensive part of the online calibration process---is parallelized.

\section{Conclusions}
\label{sec:Conclusions}
This paper addressed several issues pertaining to the operationalization of EKF-based solution approaches for the online calibration of real-time DTA systems, especially for congested scenarios and large-scale road networks. First, the concept of state augmentation is revisited to handle violations of the Markovian assumption typically implicit in online applications of the EKF. Further, we demonstrate its implementation using a real-time DTA system and analyze the computational cost of applying this technique. Second, a method based on graph-ing is proposed to operationalize the partitioned finite-difference approach that enhances scalability of the gradient computations. Synthetic experiments and a real world case study demonstrate that application of the proposed approaches yields improvements in terms of both prediction accuracy and computational performance. 

Avenues for future research include examining the performance and robustness of the approaches for the simultaneous calibration of demand and supply parameters. Further, the performance of the PSP-EKF algorithm in the presence of non-recurrent events and incidents, and estimation errors in the AR process also warrants more systematic investigation.  

\section*{Acknowledgement}
This research is supported by the National Research Foundation, Prime Minister\textquoteright s Office, Singapore, under its CREATE program, Singapore-MIT Alliance for Research and Technology (SMART) Future Urban Mobility (FM) IRG.

\bibliographystyle{unsrtnat}
\bibliography{mybibfile}  

\begin{thebibliography}{49}
\providecommand{\natexlab}[1]{#1}
\providecommand{\url}[1]{\texttt{#1}}
\expandafter\ifx\csname urlstyle\endcsname\relax
  \providecommand{\doi}[1]{doi: #1}\else
  \providecommand{\doi}{doi: \begingroup \urlstyle{rm}\Url}\fi

\bibitem[{FHWA}(2017)]{series2016fhwa}
{FHWA}.
\newblock 2016 urban congestion trends: Using technology to measure, manage,
  and improve operations.
\newblock Technical report, Federal Highway Administration, 2017.

\bibitem[{FHWA}(2010)]{taylor20102009}
{FHWA}.
\newblock 2009 urban congestion trends: How operations is solving congestion
  problems.
\newblock Technical report, Federal Highway Administration, 2010.

\bibitem[Schrank et~al.(2015)Schrank, Eisele, Lomax, and Bak]{schrank2015urban}
David Schrank, Bill Eisele, Tim Lomax, and Jim Bak.
\newblock 2015 urban mobility scorecard.
\newblock 2015.

\bibitem[Cascetta(1984)]{cascetta1984estimation}
Ennio Cascetta.
\newblock Estimation of trip matrices from traffic counts and survey data: a
  generalized least squares estimator.
\newblock \emph{Transportation Research Part B: Methodological}, 18\penalty0
  (4-5):\penalty0 289--299, 1984.

\bibitem[Bell(1983)]{bell1983estimation}
Michael~GH Bell.
\newblock The estimation of an origin-destination matrix from traffic counts.
\newblock \emph{Transportation Science}, 17\penalty0 (2):\penalty0 198--217,
  1983.

\bibitem[Maher(1983)]{maher1983inferences}
MJ~Maher.
\newblock Inferences on trip matrices from observations on link volumes: a
  bayesian statistical approach.
\newblock \emph{Transportation Research Part B: Methodological}, 17\penalty0
  (6):\penalty0 435--447, 1983.

\bibitem[Cascetta et~al.(1993)Cascetta, Inaudi, and
  Marquis]{cascetta1993dynamic}
Ennio Cascetta, Domenico Inaudi, and Gerald Marquis.
\newblock Dynamic estimators of origin-destination matrices using traffic
  counts.
\newblock \emph{Transportation Science}, 27\penalty0 (4):\penalty0 363--373,
  1993.

\bibitem[Ashok(1996)]{ashok1996estimation}
Kalidas Ashok.
\newblock \emph{Estimation and prediction of time-dependent origin-destination
  flows}.
\newblock PhD thesis, Massachusetts Institute of Technology, 1996.
\newblock URL \url{http://dspace.mit.edu/}.

\bibitem[Okutani and Stephanedes(1984)]{okutani1984dynamic}
Iwao Okutani and Yorgos~J Stephanedes.
\newblock Dynamic prediction of traffic volume through {Kalman} filtering
  theory.
\newblock \emph{Transportation Research Part B: Methodological}, 18\penalty0
  (1):\penalty0 1--11, 1984.

\bibitem[Cascetta et~al.(2013)Cascetta, Papola, Marzano, Simonelli, and
  Vitiello]{cascetta2013quasi}
Ennio Cascetta, Andrea Papola, Vittorio Marzano, Fulvio Simonelli, and Iolanda
  Vitiello.
\newblock Quasi-dynamic estimation of o--d flows from traffic counts:
  Formulation, statistical validation and performance analysis on real data.
\newblock \emph{Transportation Research Part B: Methodological}, 55:\penalty0
  171--187, 2013.

\bibitem[Osorio(2019)]{osorio2019high}
Carolina Osorio.
\newblock High-dimensional offline origin-destination (od) demand calibration
  for stochastic traffic simulators of large-scale road networks.
\newblock \emph{Transportation Research Part B: Methodological}, 124:\penalty0
  18--43, 2019.

\bibitem[Balakrishna et~al.(2007)Balakrishna, Ben-Akiva, and
  Koutsopoulos]{balakrishna2007offline}
Ramachandran Balakrishna, Moshe Ben-Akiva, and Haris~N Koutsopoulos.
\newblock Offline calibration of dynamic traffic assignment: simultaneous
  demand-and-supply estimation.
\newblock \emph{Transportation Research Record}, 2003\penalty0 (1):\penalty0
  50--58, 2007.

\bibitem[Lu et~al.(2015{\natexlab{a}})Lu, Xu, Antoniou, and
  Ben-Akiva]{lu2015enhanced}
Lu~Lu, Yan Xu, Constantinos Antoniou, and Moshe Ben-Akiva.
\newblock An enhanced {SPSA} algorithm for the calibration of dynamic traffic
  assignment models.
\newblock \emph{Transportation Research Part C: Emerging Technologies},
  51:\penalty0 149--166, 2015{\natexlab{a}}.

\bibitem[Tympakianaki et~al.(2015)Tympakianaki, Koutsopoulos, and
  Jenelius]{tympakianaki2015c}
Athina Tympakianaki, Haris~N Koutsopoulos, and Erik Jenelius.
\newblock {C-SPSA}: Cluster-wise simultaneous perturbation stochastic
  approximation algorithm and its application to dynamic origin--destination
  matrix estimation.
\newblock \emph{Transportation Research Part C: Emerging Technologies},
  55:\penalty0 231--245, 2015.

\bibitem[Oh et~al.(2019)Oh, Seshadri, Azevedo, and Ben-Akiva]{oh2019demand}
Simon Oh, Ravi Seshadri, Carlos~Lima Azevedo, and Moshe~E Ben-Akiva.
\newblock Demand calibration of multimodal microscopic traffic simulation using
  weighted discrete {SPSA}.
\newblock \emph{Transportation Research Record}, page 0361198119842107, 2019.

\bibitem[Qurashi et~al.(2019)Qurashi, Ma, Chaniotakis, and
  Antoniou]{qurashi2019pc}
Moeid Qurashi, Tao Ma, Emmanouil Chaniotakis, and Constantinos Antoniou.
\newblock {PC-SPSA}: Employing dimensionality reduction to limit {SPSA} search
  noise in {DTA} model calibration.
\newblock \emph{IEEE Transactions on Intelligent Transportation Systems}, 2019.

\bibitem[Cipriani et~al.(2011)Cipriani, Florian, Mahut, and
  Nigro]{cipriani2011gradient}
Ernesto Cipriani, Michael Florian, Michael Mahut, and Marialisa Nigro.
\newblock A gradient approximation approach for adjusting temporal
  origin--destination matrices.
\newblock \emph{Transportation Research Part C: Emerging Technologies},
  19\penalty0 (2):\penalty0 270--282, 2011.

\bibitem[Djukic(2014)]{djukic2014dynamic}
Tamara Djukic.
\newblock Dynamic od demand estimation and prediction for dynamic traffic
  management.
\newblock 2014.

\bibitem[Ashok and Ben-Akiva(2000)]{ashok2000alternative}
Kalidas Ashok and Moshe~E Ben-Akiva.
\newblock Alternative approaches for real-time estimation and prediction of
  time-dependent origin--destination flows.
\newblock \emph{Transportation Science}, 34\penalty0 (1):\penalty0 21--36,
  2000.

\bibitem[Zhou and Mahmassani(2007)]{zhou2007structural}
Xuesong Zhou and Hani~S Mahmassani.
\newblock A structural state space model for real-time traffic
  origin--destination demand estimation and prediction in a day-to-day learning
  framework.
\newblock \emph{Transportation Research Part B: Methodological}, 41\penalty0
  (8):\penalty0 823--840, 2007.

\bibitem[Bierlaire and Crittin(2004)]{bierlaire2004efficient}
Michel Bierlaire and Frank Crittin.
\newblock An efficient algorithm for real-time estimation and prediction of
  dynamic od tables.
\newblock \emph{Operations Research}, 52\penalty0 (1):\penalty0 116--127, 2004.

\bibitem[Cantelmo et~al.(2015)Cantelmo, Viti, Cipriani, and
  Nigro]{cantelmo2015improving}
Guido Cantelmo, Francesco Viti, Ernesto Cipriani, and Marialisa Nigro.
\newblock Improving the reliability of a two-steps dynamic demand estimation
  approach by sequentially adjusting generations and distributions.
\newblock 2015.

\bibitem[Zhang et~al.(2017)Zhang, Seshadri, Prakash, Pereira, Antoniou, and
  Ben-Akiva]{zhang2017improved}
Haizheng Zhang, Ravi Seshadri, A~Arun Prakash, Francisco~C Pereira,
  Constantinos Antoniou, and Moshe~E Ben-Akiva.
\newblock Improved calibration method for dynamic traffic assignment models:
  Constrained extended {Kalman} filter.
\newblock \emph{Transportation Research Record: Journal of the Transportation
  Research Board}, \penalty0 (2667):\penalty0 142--153, 2017.

\bibitem[Ashok and Ben-Akiva(2002)]{ashok2002estimation}
Kalidas Ashok and Moshe~E Ben-Akiva.
\newblock Estimation and prediction of time-dependent origin-destination flows
  with a stochastic mapping to path flows and link flows.
\newblock \emph{Transportation Science}, 36\penalty0 (2):\penalty0 184--198,
  2002.

\bibitem[Marzano et~al.(2018)Marzano, Papola, Simonelli, and
  Papageorgiou]{marzano2018kalman}
Vittorio Marzano, Andrea Papola, Fulvio Simonelli, and Markos Papageorgiou.
\newblock A kalman filter for quasi-dynamic od flow estimation/updating.
\newblock \emph{IEEE Transactions on Intelligent Transportation Systems},
  19\penalty0 (11):\penalty0 3604--3612, 2018.

\bibitem[Cantelmo et~al.(2020)Cantelmo, Qurashi, Prakash, Antoniou, and
  Viti]{cantelmo2020incorporating}
Guido Cantelmo, Moeid Qurashi, A~Arun Prakash, Constantinos Antoniou, and
  Francesco Viti.
\newblock Incorporating trip chaining within online demand estimation.
\newblock \emph{Transportation Research Part B: Methodological}, 132, 2020.

\bibitem[Barcel{\'o} et~al.(2010)Barcel{\'o}, Montero, Marqu{\'e}s, and
  Carmona]{barcelo2010travel}
Jaume Barcel{\'o}, Lidin Montero, Laura Marqu{\'e}s, and Carlos Carmona.
\newblock Travel time forecasting and dynamic origin-destination estimation for
  freeways based on bluetooth traffic monitoring.
\newblock \emph{Transportation research record}, 2175\penalty0 (1):\penalty0
  19--27, 2010.

\bibitem[Barcel{\'o} et~al.(2013)Barcel{\'o}, Montero, Bullejos, Serch, and
  Carmona]{barcelo2013kalman}
Jaume Barcel{\'o}, L{\'\i}dia Montero, Manuel Bullejos, Oriol Serch, and Carlos
  Carmona.
\newblock A kalman filter approach for exploiting bluetooth traffic data when
  estimating time-dependent od matrices.
\newblock \emph{Journal of Intelligent Transportation Systems}, 17\penalty0
  (2):\penalty0 123--141, 2013.

\bibitem[Lu et~al.(2015{\natexlab{b}})Lu, Rao, Wu, Guo, and Xia]{lu2015kalman}
Zhenbo Lu, Wenming Rao, Yao-Jan Wu, Li~Guo, and Jingxin Xia.
\newblock A kalman filter approach to dynamic od flow estimation for urban road
  networks using multi-sensor data.
\newblock \emph{Journal of Advanced Transportation}, 49\penalty0 (2):\penalty0
  210--227, 2015{\natexlab{b}}.

\bibitem[Zhou and Mahmassani(2002)]{zhou2002dynamic}
Xuesong Zhou and Hani~S Mahmassani.
\newblock Dynamic programming approach for online freeway flow propagation
  adjustment.
\newblock \emph{Transportation Research Record}, 1802\penalty0 (1):\penalty0
  263--270, 2002.

\bibitem[Antoniou et~al.(2007)Antoniou, Ben-Akiva, and
  Koutsopoulos]{antoniou2007nonlinear}
Constantinos Antoniou, Moshe Ben-Akiva, and Haris~N Koutsopoulos.
\newblock Nonlinear {Kalman} filtering algorithms for on-line calibration of
  dynamic traffic assignment models.
\newblock \emph{IEEE Transactions on Intelligent Transportation Systems},
  8\penalty0 (4):\penalty0 661--670, 2007.

\bibitem[Antoniou(2004)]{antoniou2004line}
Constantinos Antoniou.
\newblock \emph{On-line calibration for dynamic traffic assignment}.
\newblock PhD thesis, Massachusetts Institute of Technology, 2004.
\newblock URL \url{http://dspace.mit.edu/}.

\bibitem[Hashemi and Abdelghany(2015)]{hashemi2015integrated}
Hossein Hashemi and Khaled Abdelghany.
\newblock Integrated method for online calibration of real-time traffic network
  management systems.
\newblock \emph{Transportation Research Record}, 2528\penalty0 (1):\penalty0
  106--115, 2015.

\bibitem[Zhang et~al.(2018)Zhang, Seshadri, Prakash, Antoniou, Pereira, and
  Ben-Akiva]{zhang2018towards}
Haizheng Zhang, Ravi Seshadri, A~Arun Prakash, Constantinos Antoniou,
  Francisco~Camara Pereira, and Moshe Ben-Akiva.
\newblock Towards dynamic bayesian networks: State augmentation for online
  calibration of {DTA} systems.
\newblock In \emph{2018 21st International Conference on Intelligent
  Transportation Systems (ITSC)}, pages 1745--1750. IEEE, 2018.

\bibitem[Djukic et~al.(2012)Djukic, Van~Lint, and
  Hoogendoorn]{djukic2012application}
Tamara Djukic, JWC Van~Lint, and SP~Hoogendoorn.
\newblock Application of principal component analysis to predict dynamic
  origin--destination matrices.
\newblock \emph{Transportation Research Record}, 2283\penalty0 (1):\penalty0
  81--89, 2012.

\bibitem[Prakash et~al.(2017)Prakash, Seshadri, Antoniou, Pereira, and
  Ben-Akiva]{prakash2017reducing}
A~Arun Prakash, Ravi Seshadri, Constantinos Antoniou, Francisco~C Pereira, and
  Moshe~E Ben-Akiva.
\newblock Reducing the dimension of online calibration in dynamic traffic
  assignment systems.
\newblock \emph{Transportation Research Record}, 2667\penalty0 (1):\penalty0
  96--107, 2017.

\bibitem[Prakash et~al.(2018)Prakash, Seshadri, Antoniou, Pereira, and
  Ben-Akiva]{prakash2018improving}
A~Arun Prakash, Ravi Seshadri, Constantinos Antoniou, Francisco~C Pereira, and
  Moshe Ben-Akiva.
\newblock Improving scalability of generic online calibration for real-time
  dynamic traffic assignment systems.
\newblock \emph{Transportation Research Record}, 2672\penalty0 (48):\penalty0
  79--92, 2018.

\bibitem[Frederix et~al.(2014)Frederix, Viti, Himpe, and
  Tamp{\`e}re]{frederix2014dynamic}
Rodric Frederix, Francesco Viti, Willem~WE Himpe, and Chris~MJ Tamp{\`e}re.
\newblock Dynamic origin--destination matrix estimation on large-scale
  congested networks using a hierarchical decomposition scheme.
\newblock \emph{Journal of Intelligent Transportation Systems}, 18\penalty0
  (1):\penalty0 51--66, 2014.

\bibitem[Huang(2010)]{huang2010algorithmic}
Enyang Huang.
\newblock Algorithmic and implementation aspects of on-line calibration of
  dynamic traffic assignment.
\newblock Master's thesis, Massachusetts Institute of Technology, 2010.
\newblock URL \url{http://dspace.mit.edu/}.

\bibitem[Zhang(2018)]{zhang2018online}
Haizheng Zhang.
\newblock \emph{Online calibration for simulation-based dynamic traffic
  assignment: towards large-scale and real-time performance}.
\newblock PhD thesis, Massachusetts Institute of Technology, 2018.

\bibitem[Peeta and Ziliaskopoulos(2001)]{peeta2001foundations}
Srinivas Peeta and Athanasios~K Ziliaskopoulos.
\newblock Foundations of dynamic traffic assignment: The past, the present and
  the future.
\newblock \emph{Networks and Spatial Economics}, 1\penalty0 (3-4):\penalty0
  233--265, 2001.

\bibitem[Coleman and Mor{\'e}(1983)]{coleman1983estimation}
Thomas~F Coleman and Jorge~J Mor{\'e}.
\newblock Estimation of sparse jacobian matrices and graph coloring problems.
\newblock \emph{SIAM journal on Numerical Analysis}, 20\penalty0 (1):\penalty0
  187--209, 1983.

\bibitem[Castillo et~al.(2015)Castillo, Grande, Calvi{\~n}o, Szeto, and
  Lo]{castillo2015state}
Enrique Castillo, Zacar{\'\i}as Grande, Aida Calvi{\~n}o, Wai~Yuen Szeto, and
  Hong~K Lo.
\newblock A state-of-the-art review of the sensor location, flow observability,
  estimation, and prediction problems in traffic networks.
\newblock \emph{Journal of Sensors}, 2015, 2015.

\bibitem[Castillo et~al.(2008)Castillo, Conejo, Men{\'e}ndez, and
  Jim{\'e}nez]{castillo2008observability}
Enrique Castillo, Antonio~J Conejo, Jos{\'e}~Mar{\'\i}a Men{\'e}ndez, and Pilar
  Jim{\'e}nez.
\newblock The observability problem in traffic network models.
\newblock \emph{Computer-Aided Civil and Infrastructure Engineering},
  23\penalty0 (3):\penalty0 208--222, 2008.

\bibitem[Yang et~al.(2018)Yang, Fan, and Wets]{yang2018stochastic}
Yudi Yang, Yueyue Fan, and Roger~JB Wets.
\newblock Stochastic travel demand estimation: Improving network
  identifiability using multi-day observation sets.
\newblock \emph{Transportation Research Part B: Methodological}, 107:\penalty0
  192--211, 2018.

\bibitem[Ehlert et~al.(2006)Ehlert, Bell, and Grosso]{ehlert2006optimisation}
Anett Ehlert, Michael~GH Bell, and Sergio Grosso.
\newblock The optimisation of traffic count locations in road networks.
\newblock \emph{Transportation Research Part B: Methodological}, 40\penalty0
  (6):\penalty0 460--479, 2006.

\bibitem[Wen(2008)]{wen2008scalability}
Yang Wen.
\newblock \emph{Scalability of dynamic traffic assignment}.
\newblock PhD thesis, Massachusetts Institute of Technology, 2008.

\bibitem[{Google Maps}(2020)]{googlemap2020sgp}
{Google Maps}.
\newblock Singapore road network, 2020.
\newblock URL \url{http://www.google.com/maps/@1.3482868,103.756796,12z}.
\newblock [Online; accessed Oct 19, 2020].

\bibitem[Sorenson(1970)]{sorenson1970least}
Harold~W Sorenson.
\newblock Least-squares estimation: from {Gauss} to {Kalman}.
\newblock \emph{IEEE Spectrum}, 7\penalty0 (7):\penalty0 63--68, 1970.

\end{thebibliography}






\end{document}